\documentclass[journal]{IEEEtran}
\ifCLASSINFOpdf
\else
\fi
\usepackage{booktabs} 

\usepackage[english]{babel}
\usepackage{moresize}
\usepackage{amsmath}

\usepackage{algorithmic} 
\usepackage{balance}
\usepackage{multirow}
\usepackage{comment}
\usepackage{paralist}

\usepackage[english]{babel}
\usepackage[latin1]{inputenc}
\usepackage{mathrsfs}
\usepackage{graphicx}
\usepackage{amssymb}
\usepackage{url}
\usepackage{mathrsfs}
\usepackage{subfigure}
\usepackage{amsmath}
\usepackage{enumitem}
\usepackage[linesnumbered,algoruled,boxed,lined]{algorithm2e}
\usepackage{adjustbox}
\usepackage[justification=centering]{caption}

\usepackage{pgfplots}
\usetikzlibrary{pgfplots.dateplot}

\newcommand{\hide}[1]{} 

\newcommand{\etal}{\textit{et al}.}

\newcommand{\ie}{\textit{i}.\textit{e}.}
\newcommand{\eg}{\textit{e}.\textit{g}.} 
\newcommand{\wrt}{\textit{w}.\textit{r}.\textit{t}}

\def\model{MBRec}

\begin{document}
%
\title{Multi-Behavior Graph Neural Networks for Recommender System}
%
%
%

\author{Lianghao Xia,
	Chao Huang$^*$,
	Yong Xu, \textit{Senior Member, IEEE},
	Peng Dai,
	Liefeng Bo
	\thanks{L. Xia, C. Huang are with the Department of Computer Science \& Musketeers Foundation Institute of Data Science, University of Hong Kong, Hong Kong, China. E-mail: Email: aka\_xia@foxmail.com, chaohuang75@gmail.com.}
	\thanks{Y. Xu is with the School of Computer Science and Technology, South China University of Technology, Guangzhou 510006, China. Email: yxu@scut.edu.cn.}
	\thanks{P. Dai, L. Bo are affiliated with JD silicon valley research center, 675 E Middlefield Rd, Mountain View, CA 94043, USA. Email: \{peng.dai, liefeng.bo\}@jd.com.\\
    \IEEEcompsocthanksitem *Chao Huang is the corresponding author.}
}

\maketitle

\begin{abstract}
Recommender systems have been demonstrated to be effective to meet user's personalized interests for many online services (\eg, E-commerce and online advertising platforms). Recent years have witnessed the emerging success of many deep learning-based recommendation models for augmenting collaborative filtering architectures with various neural network architectures, such as multi-layer perceptron and autoencoder. However, the majority of them model the user-item relationship with single type of interaction, while overlooking the diversity of user behaviors on interacting with items, which can be click, add-to-cart, tag-as-favorite and purchase. Such various types of interaction behaviors have great potential in providing rich information for understanding the user preferences. In this paper, we pay special attention on user-item relationships with the exploration of multi-typed user behaviors. Technically, we contribute a new multi-behavior graph neural network (\model), which specially accounts for diverse interaction patterns as well as the underlying cross-type behavior inter-dependencies. In the \model\ framework, we develop a graph-structured learning framework to perform expressive modeling of high-order connectivity in behavior-aware user-item interaction graph. After that, a mutual relation encoder is proposed to adaptively uncover complex relational structures and make aggregations across layer-specific behavior representations. Through comprehensive evaluation on real-world datasets, the advantages of our \model\ method have been validated under different experimental settings. Further analysis verifies the positive effects of incorporating the multi-behavioral context into the recommendation paradigm. Additionally, the conducted case studies offer insights into the interpretability of user multi-behavior representations. We release our model implementation at https://github.com/akaxlh/MBRec.
\end{abstract}


%
\IEEEpeerreviewmaketitle

\section{Introduction}
\label{sec:intro}

With the growth of Internet services and mobile applications, recommender systems have played an increasingly critical role in addressing the information overload for many online platforms~\cite{zhao2018recommendations,han2019adaptive,yang2022knowledge}. For example, the benefits of recommendation systems could lie in providing personalized recommendations in e-commerce sites (\eg, Amazon and Taobao), or satisfying users' interest in online music streaming services (\eg, Pandora and Spotify). Currently, collaborative filtering techniques serve as one of the most important paradigms to accurately understand the preferences of users, based on their interaction behaviors~\cite{zheng2018spectral,shi2018heterogeneous}.

With the remarkable success of deep learning, there exist renewed interests in modeling user-item interactions with various neural network architectures, such as multi-layer perceptron~\cite{he2017neuralncf,sheu2021knowledge}, autoencoder network~\cite{sedhain2015autorec} and neural autoregressive models~\cite{zheng2016neural}. Built on the recent strength of graph neural networks, several studies seek to aggregate feature information from the graph-structured relational data generated by the observed user behaviors~\cite{wang2019neural,zhang2019star}. These neural network models generally focus on single type of user interaction behaviors over items, during the vectorized representation procedure of users and items. However, in real-life applications, items are often interacted by users with diverse ways~\cite{gao2019neural,zhang2020multiplex,yu2022multiplex}. For example, users can view, tag-as-favourite and purchase different products in E-commerce platforms. In such real-life scenarios, effectively modeling of multi-typed user-item interactions can provide auxiliary knowledge to characterize the diverse user behavior semantics for interest representation in recommender systems~\cite{guo2019buying,wei2022contrastive}.
For simplifying the model design, the embedding functions in most existing recommendation models nearly ignore the explicit encoding of multi-behavior collaborative signals, which are insufficient to yield satisfactory latent representations for both users and items. In this paper, we tackle the multi-behavior recommendation by enhancing user preference learning with the exploration of multi-typed user behavior data.

\begin{figure}[t]
    \centering
    \includegraphics[width=0.48\textwidth]{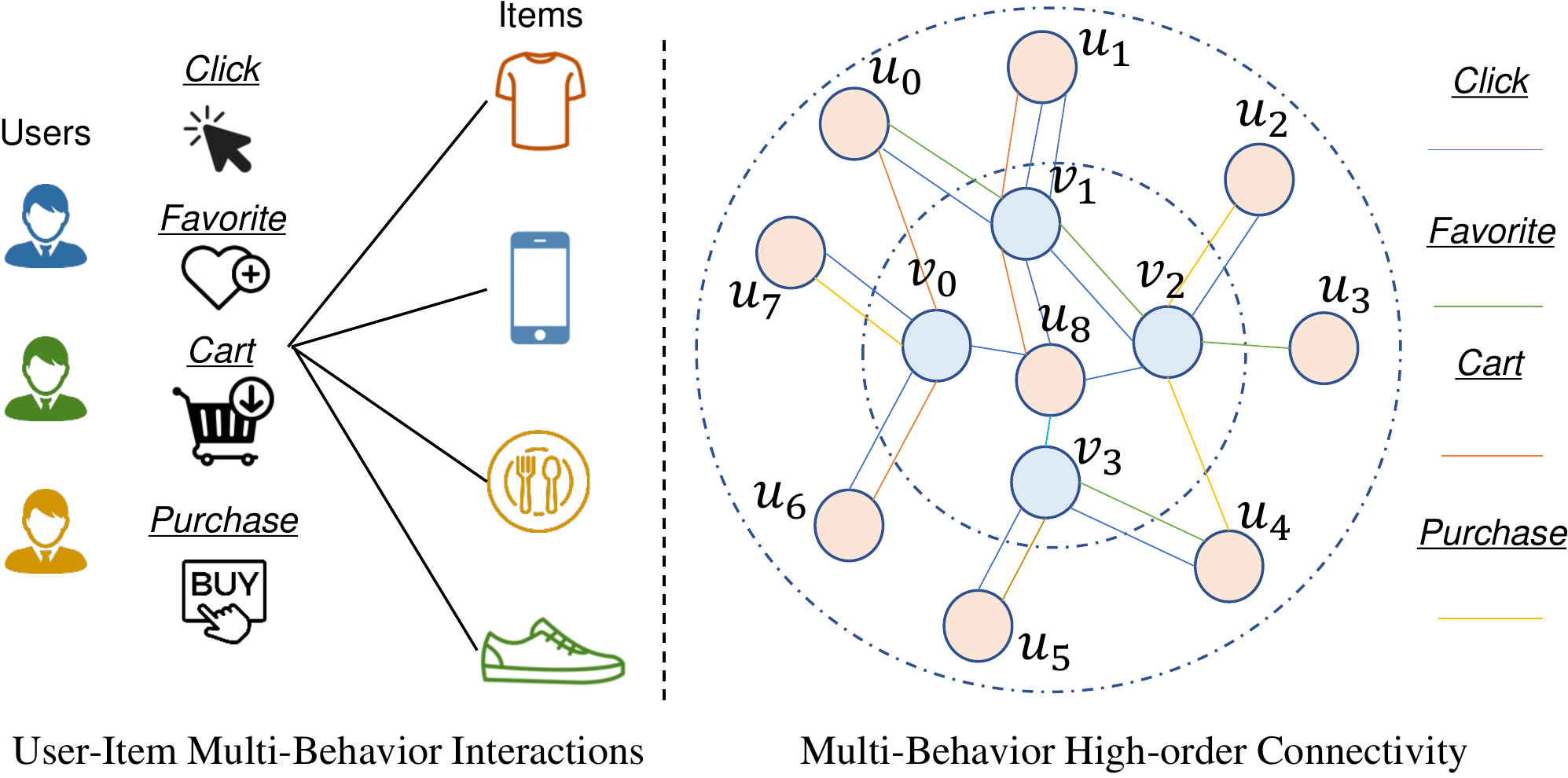}
    \vspace{-0.05in}
    \caption{Illustration of the user-item multi-behavior interactions and the corresponding multi-behavior high-order connectivity. Best viewed in color.}
    \label{fig:intro}
    \vspace{-0.1in}
\end{figure}

Although it is desirable to consider the behavior diversity in user interest representation learning for accurate recommendations, it is not a trivial task to capture the complex multi-behavioral collaborative relations. In particular, each type of user behaviors has its own interaction contexts and there exist complex dependencies across various types of interactions. Different behavior views usually provide complementary information for encoding user's interests. Therefore, in order to learn meaningful behavior representations from multi-typed user-item interactions, an effective cross-type behavior dependency modeling is a necessity in solving the multi-behavior recommendation. In addition, in the view of user-item interaction graph, the effectiveness of exploring subgraph structures has been shown in recently emerged graph-based methods (\eg, PinSage~\cite{ying2018graph} and NGCF~\cite{wang2019neural}), with the consideration of high-hop neighbors. However, to design effective embedding function in the recommendation architecture for representation aggregations, it is crucial to expressively model the high-order multi-behavior patterns of users over the interaction graph structure from different propagation layers. We show the multi-behavior recommendation scenario with the illustrated examples in Figure~\ref{fig:intro}. We can observe that user can interact with items with different behavior types (differentiated with weight lines), \eg, click, add-to-cart and purchase. In such cases, we generate a multi-behavior interaction graph to represent the diverse collaborative relations among users and items with the high-order multiplex connectivity information. \\\vspace{-0.1in}



\noindent \textbf{Present Work.} Motivated by the aforementioned challenges, this work proposes a new recommendation framework: \underline{M}ulti-\underline{B}ehavior \underline{G}raph \underline{N}eural \underline{N}etwork (\model) that explicitly incorporates multi-typed behavior context into the encoding of diverse user preference. In the proposed \model\ framework, we can capture the heterogeneous relationships across different types of user-item interactions. Specifically, to cope with the cross-type behavior dependencies, we propose an attention-enhanced graph neural network to preserve the high-order interaction patterns over the multiplex graph structures. Through this way, \model\ is able to preserve the fine-grained semantics of user-item relations and facilitate the modeling of diverse user interest.
To differentiate the influences of different types of behaviors, a gated aggregation mechanism is developed to help fuse the contextual signals from different types of behaviors for a better embedding learning. In addition, we endow our \model\ with the capability of aggregating high-order behavior representations in an adaptive way. To achieve this goal, a mutual relation encoder is proposed to learn summarized representations across different graph layers. This component allows our model to better capture and interpret the global property of user-item interaction graph. 

Lastly, it is worth mentioning that although the multi-behavior information has been considered in recent studies~\cite{gao2019neural,gao2019learning}. these work only consider the multi-behavior dependencies with the predefined correlations, which can hardly comprehensively capture the complex cross-type behavior dependencies in real-world recommendation scenarios. In addition, another study of multi-behavior recommendation model proposes to consider the dependencies between different interactions based on multi-channel graph convolutional network~\cite{jin2020multi}. Different from these approaches, we contribute a new recommendation framework to explicitly exploit the high-order collaborative signals in the form of multi-behavior patterns. To enhance the global multi-behavior relation learning, we design a new graph aggregation scheme with order-wise mutual dependency modeling, which automatically differentiates the importance of behavior-aware representations from different graph hops during the message passing process. Lack of considering the complex global multi-behavior patterns could easily lead to suboptimal representations for user preference modeling.


In summary, we highlight our contributions as follows:

\begin{itemize}[leftmargin=*]

\item We introduce a new multi-behavior recommendation framework, which explores the high-order and cross-type behavior inter-dependencies in a hierarchical manner.

\item We propose a new recommendation framework that inherits the merits of graph-enhanced collaborative filtering paradigm and designs a multi-behavior propagation strategy for heterogeneous user-item interactions.

\item Then, a graph-structured mutual relation encoder is developed to explore high-order user preference and promote the collaboration of different layer-specific patterns for robust multi-behavior representations. Furthermore, a graph sampling algorithm is developed to improve the scalability of \model\ for dealing with large-scale graph data.

\item Experimental results on three real-world datasets show the superiority of our \model\ model over a variety of baselines. We further perform the model ablation study to better understand the effect of our designed sub-modules. 


\end{itemize}

In this paper, we propose to advance our previous work~\cite{icdegnmr} from the following aspects: i) Different from our previous method which overlooks the cross-layer implicit dependency between behavior representations, we propose a new recommendation framework to explicitly promote the cooperation of behavior patterns across different graph layers, for more accurate user and item representations (Section~\ref{sec:solution}). ii) We provide a comprehensive complexity analysis and computational cost evaluation of our model to show that our proposed model could achieve competitive time efficiency as compared to most state-of-the-art techniques (Section~\ref{sec:complexity} and Section~\ref{sec:time_eval}). iii) We further evaluate the model performance with respect to different sparsity levels of user-item interaction data, to justify the robustness of our multi-behavior recommendation method in capturing user's preference under different data sparsity degrees (Section~\ref{sec:sparsity}). iv) We add three new recently developed baselines (\ie, MBGCN, MATN and NGCF+M) in our performance evaluation to show the superiority of our method (Section~\ref{sec:eval}). Additionally, we present more hyperparameter study results with respect to the hidden state dimensionality and behavior embedding channels (Section~\ref{sec:hyper_study}). v) We perform case studies to show the interpretation capability of our approach in capturing the behavior relationships as well as the layer-wise representation dependence (Section~\ref{subsec:case}). vi) We adopt two new datasets (BeiBei and IJCAI-contest) collected from real-world e-commerce platforms for performance evaluation across different experimental settings. The user-item interactions in online retailing systems are multiplex in nature and can well reflect the relation heterogeneity between users and items. vii) Finally, we present detailed discussion about the related work from three research lines: neural collaborative filtering techniques, recommendation with multi-behavior modeling and graph neural networks (Section~\ref{sec:relate}). \\\vspace{-0.1in}


\section{Preliminaries}
\label{sec:model}


We begin by describing the multi-behavior recommendation and introducing key notations. Suppose we have a recommendation scenario with a set of users $U$ ($u_i\in U$) and a set of items $V$ ($v_j\in V$). Here, the index of user and item is denoted by $i$ ($i\in [1,...,I]$) and $j$ ($j\in [1,...,J]$), respectively. Different from most existing recommender systems which associate users with their interacted items based on singular type of user-item relations, this work explores the inter-dependent relations across different types of user-item behaviors (\eg, click, tag-as-favorite, review, like, or purchase). \\\vspace{-0.12in}

\noindent Definition 1. \textbf{Multi-Behavior Interaction Tensor} $\textbf{X}$. To represent the multi-typed interactions between users and items, we define a three-way multi-behavior interaction tensor $\textbf{X} \in \mathbb{R}^{I\times J\times K}$, where $K$ (indexed by $k$) denotes the number of types of user-item interactions. Given the $k$-th type of interactions, the corresponding element $x_{i,j}^k \in \textbf{X}$ is set to 1 if the item $v_j$ has been adopted by user $u_i$. Otherwise, $x_{i,j}^k=0$.\\\vspace{-0.12in}


\noindent \textbf{Multi-Behavior Recommendation}. In the recommendation scenario with multi-type interaction behaviors, we first define the target behavior type (\eg, $k$-th) as our predictive objective and consider other types ($k'\in [1,...,K] \& k' \neq k$) of interaction as auxiliary behaviors. In real-world recommendation scenarios, the target behavior type can be set according to different task-specific requirements. For example, some e-commerce systems may be more interested in ultimate purchase transactions~\cite{wu2018turning}, while forecasting click-interactive behavior is also very important for online advertising platforms~\cite{ren2018learning}. We formally present the multi-behavior recommendation as: \\\vspace{-0.12in}

\noindent \textbf{Input}: multi-behavior interaction tensor $\textbf{X} \in \mathbb{R}^{I\times J\times K}$ which jointly includes source and target behavior of users in $U$.\\
\noindent \textbf{Output}: A predictive framework which effectively forecasts the unknown target type of user-item interactions.
\section{Methodology}
\label{sec:solution}

In this section, we first elaborate the technical details of our proposed \model\ framework. Its key idea is to explore the complex inter-dependencies across different types of users' interactive behavior, to parameterize weight matrices for the relation heterogeneity aggregation, high-order message passing and propagation modules. Such process can be decomposed into two key components: (i) Multi-Behavior Graph-Structured Dependency Modeling: it jointly preserves the type-specific behavior semantics and type-wise behavior inter-dependencies within a graph-structured learning architecture. (ii) Cross-Layer Mutual Relation Learning: it captures the mutual relationships between the aggregated multi-hop feature representations of neighbors at different hops.

\begin{table}[t]
    \caption{Summary of Key Notations}
    \label{tab:data}
    \centering
    \footnotesize
	\setlength{\tabcolsep}{0.6mm}
    \begin{tabular}{c | c}
        \hline
        Notations & Description \\
        \hline
        $U$ ($u_i\in U$) & Set of users \\
        $V$ ($v_j\in V$)  & Set of items \\
        $K$ (indexed by $k$) & Number of behavior types \\
        $\textbf{X} \in \mathbb{R}^{I\times J\times K}$ & Multi-behavior interaction tensor \\
        $\textbf{H}_{i\leftarrow}^{k,(l)}$, $\textbf{H}_{j\leftarrow}^{k,(l)}$ & Behavior-aware representations of users/items \\
        $\lambda(\cdot)$ & Behavior-aware encoding function \\
        $\hat{\textbf{H}}_{i\leftarrow}^{k,(l)}$ & Recalibrated type-specific behavior embedding \\
        $\beta_k $ & Attention value for type-specific behavior \\
        $\psi(\cdot)$ & Cross-behavior message aggregation function \\
        $G=\{U,V,\mathcal{E}\}$ & Multi-behavior high-order graph \\
        $\hat{\textbf{E}}_i^{(l)}$, $\hat{\textbf{E}}_j^{(l)}$ & Cross-layer fused embedding \\
        $\mathbf{\Gamma}_{i,j}$ & Fused representation for prediction \\
        \hline
    \end{tabular}
    \label{tab:notations}
\end{table}

\begin{figure}[t]
    \centering
    \includegraphics[width=\columnwidth]{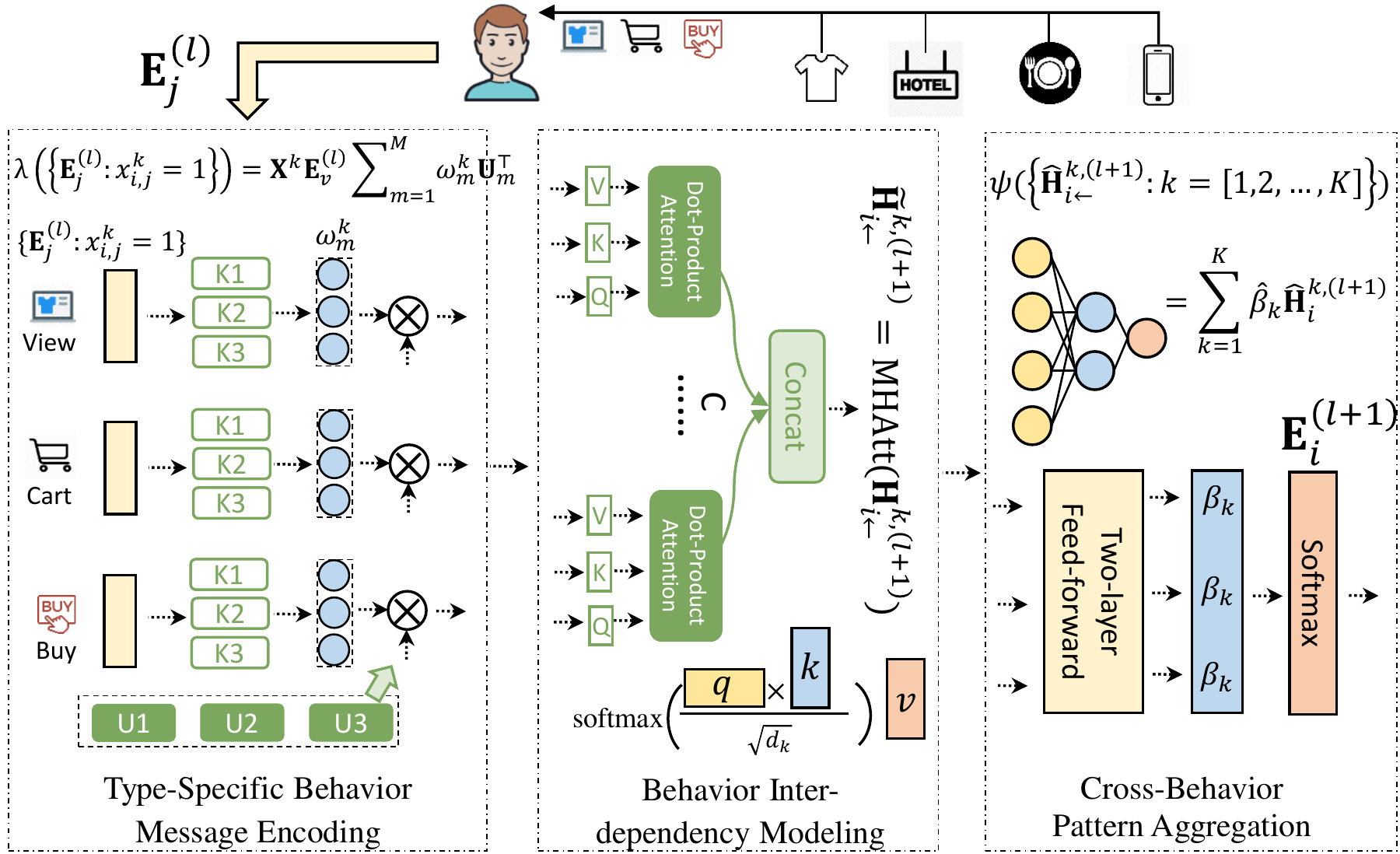}
    \caption{Multi-behavior graph dependency modeling.}
    \label{fig:framework2}
    \vspace{-0.05in}
\end{figure}

\subsection{Multi-Behavior Graph Dependency Modeling}
Figure~\ref{fig:framework2} presents the model flow of our multi-behavior graph dependency modeling. With the consideration of different types of user-item behavioral relations, we first construct the multi-behavior user-item interaction graph.\\\vspace{-0.12in}

\noindent Definition 2. \textbf{Multi-Behavior Graph} $G$. Based on the input multi-behavior interaction tensor $\textbf{X}$, we generate a graph $G=\{U,V,\mathcal{E}\}$ where user node $u_i \in U$ and item node $v_j \in V$ is connected with the edge $e_{i,j,k}\in \mathcal{E}$, if $u_i$ interacts with $v_j$ under the $k$-th behavior type (\ie, $x_{i,j}^k=1$). Each edge $e_{i,j,k}\in \mathcal{E}$ is associated with a specific behavior type of $k$. Due to the interaction heterogeneity property, there exist multiplex edges between the same user-item pair given $u_i$ interacts with $v_j$ under multiple behavior types.\\\vspace{-0.12in}

\subsubsection{\bf Behavior-aware Message Construction}
Based on the multi-behavior graph $G$, we first generate the propagated information for the user($u_i$)-item($v_j$) pair with the following behavior-aware message passing paradigm:
\begin{align}
\textbf{H}_{i\leftarrow}^{k,(l+1)} =  \lambda(\{\textbf{E}_{j}^{(l)}:x_{i,j}^k=1\})\nonumber\\
\textbf{H}_{j\leftarrow}^{k,(l+1)} =  \lambda(\{\textbf{E}_{i}^{(l)}:x_{i,j}^k=1\})
\end{align}
\noindent where $\lambda(\cdot)$ denotes the behavior-aware encoding function for preserving the semantics of individual type of interactive behavior for $u_i$ and $v_j$. Here, $\textbf{H}_{i\leftarrow}^{k,(l+1)} \in \mathbb{R}^d$ and $\textbf{H}_{j\leftarrow}^{k,(l+1)} \in \mathbb{R}^d$ with dimensionality of $d$ (output from $\lambda(\cdot)$ function) are the $(l+1)$-th layer representation which preserves the characteristics of the behavior type of $k$. In the first-layer propagation, we generate the input feature vector $\textbf{E}_{i}^{0}$ and $\textbf{E}_{j}^{0}$ of user $u_i$ and item $v_j$ with the Autoencoder-based pre-training~\cite{sedhain2015autorec} over multi-behavior interaction tensor $\textbf{X}$, to project different types of high-dimensional behavior embeddings into low-dimensional latent space. The reasons for using Autoencoder is to generate more informative initial embeddings for users and items, through the auto-encoding training paradigm.



In the multi-behavior recommendation, various types of users' behaviors reflect the preference from behavior-specific characteristics. For example, view-interactive behavior happens more frequently than add-to-cart and add-to-favorite activities~\cite{guo2019buying}. Additionally, add-to-favorite interactive behavior could provide rich information to characterize users' implicit interests over items, although the purchase behavior may be postponed and does not happen right away~\cite{cen2019representation}. Hence, we design our semantic encoding function $\lambda(\cdot)$, to capture the type-specific behavior contextual information in the message construction process.\\\vspace{-0.1in}

\noindent \textbf{Multi-Channel Behavior Embedding Layer}. Inspired by recent advancement of memory-augmented neural network models in multi-dimensional context learning~\cite{wang2018neural}, we build our behavior semantic encoder upon the multi-channel neural framework to learn customized representations for individual type of user-item interactions. Our multi-channel embedding layer utilizes the external representation units, in which each unit corresponds to a certain dimensional of behavior semantics (\eg, behavior data distributions with respect to different categories of items). Specifically, we integrate the channel-based projection layer with the behavior-aware attention mechanism to fuse the learned semantic information across different channels. For each behavior type $k$, this module is firstly equipped with a contextual transformation layer to update the input user/item embeddings $\textbf{E}_j^{(l)}$ and $\textbf{E}_j^{(l)}$ in the $l$-th graph layer. Without loss of generality, we formally present the behavior semantic encoder with $M$ (indexed by $m$) channels for the message of user $u_i$ from his/her connected item nodes $\{v_j|x_{i,j}^k=1\}$ under the behavior type of $k$ as below:
\begin{align}
    \textbf{H}_{i\leftarrow}^{k,(l+1)} &=\sum_{m=1}^M\omega_m^k\textbf{U}_m \sum_{x_{i,j}^k=1}{\textbf{E}_j^{(l)}}\nonumber\\
    \omega_m^k &=\delta(\textbf{K}\cdot \sum_{x_{i,j}^k=1}{\textbf{E}_j^{(l)}} + \textbf{b})(m)
\end{align}
\noindent where $\textbf{U}_m\in\mathbb{R}^{d\times d}$ is the $m$-th channel transformation, $\omega_m^k$ is the $m$-th weight calculated from the neighboring nodes $\{v_j|x_{i,j}^k=1\}$ of $u_i$ under behavior type of $k$. $\sum_{m=1}^M\omega_m^k\textbf{U}_m$ represents the learned behavior-type-specific contextual transformation. Additionally, $\textbf{K}\in\mathbb{R}^{M\times d}$ and $\textbf{b}\in\mathbb{R}^M$ are transformation and bias parameters to calculate the weights, and $\delta(\cdot)$ denotes ReLU activation. Similar semantic encoder can be applied to learn embedding $\textbf{H}_{j\leftarrow}^{k,(l+1)}$ between item $v_j$ and its connected users $\{u_i|x_{i,j}^k=1\}$ with the $k$-th behavior type. By performing the dimension-wise relation learning for each type of behavior patterns, the underlying semantic information can be smoothly captured with multi-channel behavior embedding in our framework.

\subsubsection{\bf Behavior Inter-dependency Modeling}
In addition to encoding the type-specific behavior semantics, another key aspect of multi-behavior relation learning lies in exploring the inter-dependencies across different types of user-item interactive behavior. For instance, add-to-cart and tag-as-favorite activities are good indicators for purchase of users. Hence, after the message propagating process with the learned representations $\textbf{H}_{i\leftarrow}^{k,(l+1)}$ and $\textbf{H}_{j\leftarrow}^{k,(l+1)}$, it is crucial to consider the inter-dependencies among different types of behaviors (\ie, $k\in [1,...,K]$), and further refine the type-specific behavior embeddings propagated from other behavior types.

Motivated by the recent success of transformer networks in distilling inter-correlations between entities~\cite{yun2019graph}, we develop a self-attentive network for multi-behavior inter-dependency modeling. Based on the paradigm of self-attention layer with scaled dot-product attention, we create three transformation matrices to project input representation into three latent dimensions, \ie, $\textbf{Q}^h\in\mathbb{R}^{\frac{d}{C}\times d}$, $\textbf{V}^h\in\mathbb{R}^{\frac{d}{C}\times d}$ and $\textbf{K}^h\in\mathbb{R}^{\frac{d}{C}\times d}$ as the query, value and key transformations for each of the $C$ (indexed by $c$) attention heads. Then, the dependency scores between the $k$-th and the $k'$-th behavior message is calculated in the dot-product manner as follows:
\begin{align}
    \alpha_{k,k'}^c &=\frac{(\textbf{Q}^c\textbf{H}^{k,(l+1)}_{i\leftarrow})^\top \cdot (\textbf{K}^c\textbf{H}^{k',(l+1)}_{i\leftarrow})}{\sqrt{\frac{d}{C}}} \nonumber\\
    \hat{\alpha}_{k,k'}^c&=\frac{\exp{\alpha_{k,k'}^c}}{\sum_{k'=1}^{k,(l+1)}\exp{\alpha_{k,k'}^c}}
\end{align}
\noindent where $\hat{\alpha}_{k,k'}^c$ is the learned quantitative correlation weight between the behavior type $k$ on $k'$, which is calculated from the intermediate score $\alpha_{k,k'}^c$ by the softmax function. To expand the model ability in capturing cross-behavior dependency from different hidden dimensions (\eg, co-occurrence frequencies and regularties), we augment the self-attention layer with the multi-head mechanism to endow the attentive relation learning under multiple represenation subspaces. In such cases, we associate each head-specific attention component with the a set of Query ($\textbf{Q}^c$), Key ($\textbf{K}^c$) and Value ($\textbf{V}^c$) weight matrices. Based on the head-specific correlation scores, we then recalibrate the type-specific behavior message by concatenating the multiple attention heads:
\begin{align}
\tilde{\textbf{H}}_{i\leftarrow}^{k,(l+1)} &= \text{MH-Att}(\textbf{H}_{i\leftarrow}^{k,(l+1)}) \nonumber\\
&=\mathop{\Bigm|\Bigm|}\limits_{c=1}^C \sum_{k'=1}^K \hat{\alpha}_{k,k'}^h\textbf{V}^c\cdot \textbf{H}_{i\leftarrow}^{k',(l+1)}
\end{align}
where $\mathop{\Bigm|\Bigm|}$ denotes the vector concatenation, and $\tilde{\textbf{H}}_{i}^{k,(l+1)}$ is the recalibrated message propagated to the target node $u_i$ under behavior type $k$. To preserve the original type-specific behavior message and prevent the gradient vanishing issue, we perform the element-wise addition between the original and recalibrated type-specific message with a residual connection scheme. That is, $\hat{\textbf{H}}_{i\leftarrow}^{k,(l+1)}=\tilde{\textbf{H}}_{i\leftarrow}^{k,(l+1)}+\textbf{H}_{i\leftarrow}^{k,(l+1)}$ is the refined type-specific behavior representation which preserves the cross-type behavior inter-dependent information.

\subsubsection{\bf Personalized Multi-Behavior Aggregation}
With the incorporation of both the type-specific behavior semantics and type-wise inter-correlations into our multi-behavior dependency modeling module, we introduce an aggregation layer to fuse interactive patterns across different types of behaviors. Formally, we define the message aggregation function as:
\begin{align}
    \textbf{E}_i^{(l+1)}=\psi(\{\hat{\textbf{H}}_{i\leftarrow}^{k,(l+1)}:k=[1,2,...,K]\})
\end{align}

In real-life online platforms (\eg, online retailing or review sites), item interactive patterns may vary by users due to their different behavior preferences. For example, some people prefer to add their interested items into the favorite list but sporadically make purchases, while the add-to-favorite action is more likely to be followed by the purchase behavior. Therefore, to aggregate message from different types of behavior embeddings and obtain expressive representations on the local user-item multi-behavior interaction graph, it is essential to identify the contribution of different types of behavior in assisting the final prediction on the target type of user behavior in a customized manner.

Towards this end, with the recalibrated type-specific behavior message $\hat{\textbf{H}}_{i\leftarrow}^{k,(l+1)}$, the multi-behavior graph encoder is further equiped with an attention network to differentiate the influences of different behavior types for user $u_i$. Specifically, the message aggregation module adopts a two-layer feed-forward network for weight estimation:
\begin{align}
    \beta_{k}&=\textbf{w}_2^\top\cdot\delta(\textbf{W}_1 \hat{\textbf{H}}_{i\leftarrow}^{k,(l+1)} + \textbf{b}_1)+b_2\nonumber\\ \hat{\beta}_k&=\frac{\exp{\beta_k}}{\sum_{k'=1}^{K}\exp{\beta_{k'}}}
\end{align}
where $\textbf{W}_1\in\mathbb{R}^{d'\times d}$, $\textbf{b}_1\in\mathbb{R}^{d'}$, $\textbf{w}_2\in\mathbb{R}^{d'}$, and $b_2\in\mathbb{R}$ are transformations and bias vectors for weights calculation, and $d'$ is the dimensionality of the intermediate hidden layer. $\delta$ is the ReLU activation function. $\beta_k$ is the intermediate attention value and $\hat{\beta}_k$ is the final influence weight of interaction type $k$ calculated using softmax function. With the calculated weights, the message aggregation layer then applies perform the weighted summation over type-specific behavior embeddings $\textbf{H}_{i\leftarrow}^{k,(l+1)}$, so as to acquire the encoded node embeddings $\textbf{E}^{(l+1)}_i=\sum_{k=1}^K\hat{\beta}_k\hat{\textbf{H}}_{i\leftarrow}^{k,(l+1)}$ corresponding to the $l$-th layer graph encoder. With the developed multi-behavior modeling component, we endow the \model\ with the capability of modeling of type-specific behavioral semantics and cross-type dependencies.

\begin{figure}
    \centering
    \includegraphics[width=0.85\columnwidth]{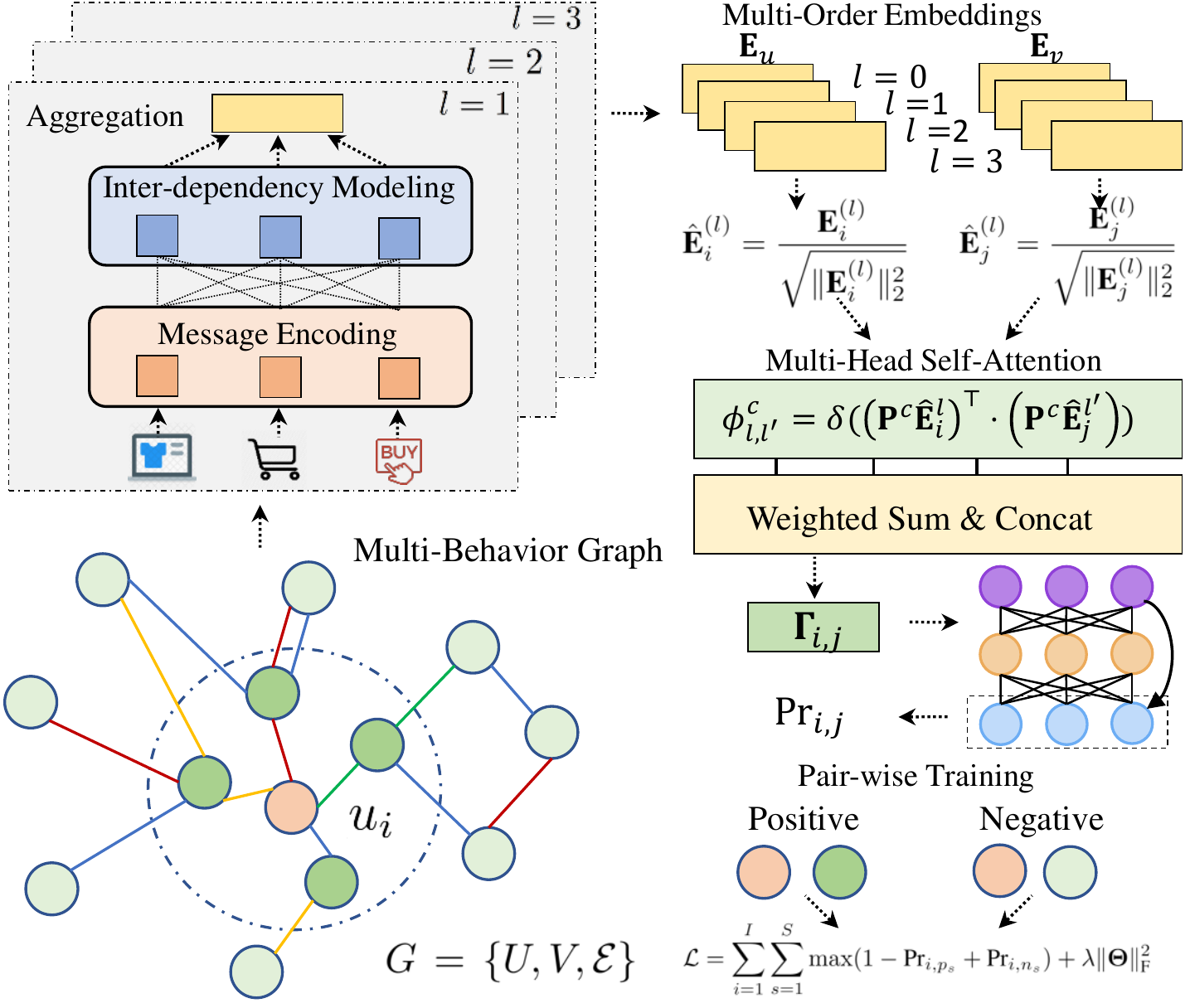}
    \vspace{-0.05in}
    \caption{High-order propagation with layer-wise mutual relation learning for multi-behavior representations.}
    \label{fig:framework1}
    \vspace{-0.2in}
\end{figure}

\subsection{High-order Mutual Relation Learning}
\subsubsection{\bf High-order Multi-Behavior Pattern Propagation}
With the multi-behavior pattern representation performed by first-order dependency learning, we can enable the high-order multi-behavior relational structures on the global graph $G=\{U,V,\mathcal{E}\}$ by stacking more embedding propagation layers (as shown in Figure~\ref{fig:framework1}). Based on our developed multi-behavior graph encoder, in the high-order information propagation process, \model\ is capable of capturing the high-order collaborative signals across different types of user-item interactive relations. After performing the multi-behavior graph encoder $l$ times, each node (\ie, $u_i$ or $v_j$) could receive the messages propagated from its neighbors with $l$-th hop distance, which is presented as follows:
\begin{align}
    \textbf{E}^{(l+1)}_{u}&=\psi(\text{MH-Att}(\lambda(\{\textbf{E}^{(l)}_j:x_{i,j}^k=1\}))\nonumber\\
    &=\sum_{k=1}^K\hat{\beta}_k\cdot\text{MH-Att}(\textbf{X}^k\textbf{E}^{(l)}_{v}\sum_{m=1}^M\omega_m^k\textbf{U}_m^\top)
\end{align}
where $\textbf{X}^k\in\mathbb{R}^{I\times J}$ is the adjacent matrix under the $k$-th behavior type, $\textbf{E}^{(l+1)}_u\in\mathbb{R}^{I\times d}$ and $\textbf{E}^{(l)}_v\in\mathbb{R}^{J\times d}$ refer to the embedding matrix of users and items, respectively.

While the high-order connectivity is exploited in the multi-behavior graph $G$, the message propagation process with the rigid order hinders the representation power in learning the global graph-structured behavior dependencies. In specific, the current graph-based message passing architecture maps the input multi-behavior user-item interactions into a fixed length embedding with the highest $L$-th layer, which results in the information loss of previous layer-specific aggregated patterns. For example, a user can interact with different items with direct (one-hop) or indirect (multi-hop) connected relationships. In such case, it is crucial to identify the most relevant signals across both low- and high-order connections, for characterizing the preference of this target user. To overcome this limitation and free the message passing graph neural architecture from fixed-length internal representation, we introduce a cross-layer mutual relation learning module, to endow our graph neural architecture \model\ with the capability to focus on certain parts of graph-structured behavioral patterns across different-order connectivity.\\\vspace{-0.12in}

\subsubsection{\bf Cross-Layer Mutual Relation Modeling} To remedy the existing shortcomings, we propose to model the importance between users' and items' multi-order embeddings with a multi-head self-attention network. For user $u_i$'s multi-order embeddings $\textbf{E}_i^{(0)},...,\textbf{E}_i^{(1)},...,\textbf{E}_i^{(L)}$ ($\textbf{E}_i^{(1)} \in\mathbb{R}^{d}$) and item $v_j$'s embeddings $\textbf{E}_j^{(0)},...,\textbf{E}_j^{(1)},...,\textbf{E}_j^{(L)}$ ($\textbf{E}_j^{(1)} \in\mathbb{R}^{d}$) learned by the stacked multi-behavioral graph encoders, we first normalize the multi-order embeddings for cross-layer embedding fusion with the following operation:
\begin{align}
\hat{\textbf{E}}_i^{(l)}=\frac{\textbf{E}_i^{(l)}}{\sqrt{\|\textbf{E}_i^{(l)}\|_2^2}};~~\hat{\textbf{E}}_j^{(l)}=\frac{\textbf{E}_j^{(l)}}{\sqrt{\|\textbf{E}_j^{(l)}\|_2^2}}
\end{align}
Based on the block of multi-head self-attention mechanism, we calculate $C$ (corresponds to $C$ heads) importance matrices $\phi^c\in\mathbb{R}^{(L+1)\times(L+1)}$ for adaptive multi-order combination. With the design of 2-D matrix $\phi^c$, we can capture the pairwise mutual relation across different layer-specific multi-behavior patterns. We present this process as follows:
\begin{align}
    \phi_{l,l'}^c=\delta((\textbf{P}^c\hat{\textbf{E}}_i^{l})^\top\cdot(\textbf{P}^c\hat{\textbf{E}}_j^{l'}))
\end{align}
where $\phi^c_{l,l'}$ is the $c$-th importance score for the combination between the $l$-th layer user embedding and $l'$-th layer item embeddings. $\textbf{P}^c\in\mathbb{R}^{\frac{d}{C}\times d}$ is the transformation to acquire key vectors, and $\delta(\cdot)$ is the ReLU activation function to learn non-linearities of feature interactions.
Based on the learned importance scores, we then generate the fused representation for prediction as follows:
\begin{align}
    \mathbf{\Gamma}_{i,j}=\mathop{\Bigm|\Bigm|}\limits_{c=1}^{C}\sum_{l=0}^L
    \sum_{l'=0}^L\phi_{l,l'}^h(\textbf{T}^c\hat{\textbf{E}}_i^{(l)})\circ(\textbf{T}^c\hat{\textbf{E}}_j^{(l')})
\end{align}
where $\circ$ denotes the element-wise production, and $\textbf{T}^c\in\mathbb{R}^{\frac{d}{C}\times d}$ is the value transformation. The $C$ head-specific representations are concatenated to generate the fused representation $\mathbf{\Gamma}_{i,j}\in\mathbb{R}^d$, which is fed into a feed-forward network to make forecasting on the unknown user-item interaction with the target behavior type $k$:
\begin{align}
\text{Pr}_{i,j}=\textbf{w}_4^\top(\delta(\textbf{W}_3\mathbf{\Gamma}_{i,j}+\textbf{b}_3)+\mathbf{\Gamma}_{i,j})
\end{align}
where $\textbf{W}_3\in\mathbb{R}^{d\times d}$, $\textbf{w}_4\in\mathbb{R}^d$ and $\textbf{b}_3\in\mathbb{R}^d$ are network parameters, $\delta(\cdot)$ is the ReLU activation. Note that a residual connection is employed for better gradients propagation.

\subsection{Graph Sampling Algorithm of \model}
\begin{algorithm}[t]
    \small
	\caption{Graph Sampling of \model\ Model}
	\label{alg:sample}
	\LinesNumbered
	\KwIn{seed users $\mathbb{U}$, seed items $\mathbb{V}$, adjacent tensor $\textbf{X}\in\mathbb{R}^{I\times J\times K}$, sampling depth $D$, sampling number per step $N$}
	\KwOut{sampled users $\hat{\mathbb{U}}$, sampled items $\hat{\mathbb{V}}$, adjacent matrix of the sampled sub-graph $\hat{\textbf{X}}$}
	Initialize the normalized adjacent matrix $\bar{\textbf{X}}\in\mathbb{R}^{I\times J}$ with $\bar{\textbf{X}}_{i,j}=\frac{\|\textbf{X}_{i,j}\|_1}{\sqrt{\|\textbf{X}_i\|_1\|\textbf{X}_j\|_1}}$ for non-zero elements\\
	Initialize sampling weights $P_u\in\mathbb{R}^I$ and $P_v\in\mathbb{R}^J$ with zeros\\
	Initialize sampled user/item set $\hat{\mathbb{U}}=\mathbb{U}$, $\hat{\mathbb{V}}=\mathbb{V}$\\
	$P_u+=\bar{\textbf{X}}_i$ for $u_i$ in $\hat{\mathbb{{U}}}$;~~ $P_v+=\bar{\textbf{X}}_j$ for $v_j$ in $\hat{\mathbb{{V}}}$\\
	
	\For{$d=1$ to $D$}{
	    $\bar{P_u}=\frac{P_u^2}{\|P_u\|_2^2}$;~~ $\bar{P_v}=\frac{P_v^2}{\|P_v\|_2^2}$\\
	    Sample $N$ unsampled users and $N$ unsampled items $\bar{\mathbb{U}}$, $\bar{\mathbb{V}}$ according to $\bar{P_u}$ and $\bar{P_v}$\\
	    $\hat{\mathbb{U}}=\hat{\mathbb{U}}\cup\bar{\mathbb{U}}$;~~ $\hat{\mathbb{V}}=\hat{\mathbb{V}}\cup\bar{\mathbb{V}}$\\
	    $P_u+=\bar{\textbf{X}}_i$ for $u_i$ in $\bar{\mathbb{U}}$;~~ $P_v+=\bar{\textbf{X}}_j$ for $v_j$ in $\bar{\mathbb{{V}}}$\\
	}
	Construct $\hat{\textbf{X}}\in\mathbb{R}^{|\hat{\mathbb{U}}|\times |\hat{\mathbb{V}}| \times K}$ by using $\textbf{X}_{i,j,k}$ for $u_i$ in $\hat{\mathbb{U}}$ and $v_j$ in $\hat{\mathbb{V}}$\\
    return $\hat{\mathbb{U}}$, $\hat{\mathbb{V}}$, $\hat{\textbf{X}}$
\end{algorithm}

In this subsection, we present the sampling algorithm of \model\ in handling large-scale user-item graph with multi-behavioral interaction information. One key limitation of graph neural architecture with full-batch mode is the requirement of performing aggregation over all nodes per graph layer. This operation involves high computational cost, which can hardly make it scalable for large-scale multi-behavior graph. To mitigate this problem and endow \model\ with the ability of serving as large-scale deep recommendation engine in real-world scenarios, we adopt graph sampling for \model\ model. 

The graph sampling algorithm is elaborated in Algorithm~\ref{alg:sample}. The key idea of our algorithm is to perform random walk-based computation graph sampling over the multi-behavior graph $G=\{U,V,\mathcal{E}\}$. Current state-of-the-art sampling strategies of graph neural models~\cite{ying2018graph,hu2020heterogeneous} largely rely on the mini-batch training paradigm over sparse local sub-graph, which may limit the model efficiency in our multi-behavior recommendation settings. The reasons are mainly two-folds: (i) While the sampled relatively small sub-graph could speed-up the model training for each step, it is likely that the user preference estimation is inaccurate if a lot of behavior-aware user-item edges are ignored during the sampling. (ii) The frequent CPU calculation and memory access also involve much computational and space cost.


To tackle the above challenges, we optimize our sampling algorithm to generate densely-connected computation graph to fit our multi-behavior recommendation scenario. In the training process of our \model\ framework, we maintain the sampling weight vectors $P_u$ and $P_v$ to contain the cumulative neighborhood information for the set of already sampled nodes. To present the dominate phenomenon of frequent users/items with large number of multi-behavior interactions, we design a normalized adjacent matrix $\bar{\textbf{X}}$ to update the sampling weights. Empirically, by sampling relatively enough node instances (\ie, $N$) for each step, the obtained sub-graphs are dense enough to contain the important multi-behavior information between users and items. Normally, each time we construct a sub-graph which contains a moderate number of nodes (\eg, tens of thousands), and the model is trained on this sampled sub-graph for one epoch (\eg, hundreds or thousands of training steps).

\subsection{Model Optimization of \model}
\begin{algorithm}[t]
    \small
	\caption{Model Optimization of \model}
	\label{alg:train}
	\LinesNumbered
	\KwIn{multi-behavior interaction tensor $\textbf{X}\in\mathbb{R}^{I\times J\times K}$, initial node embeddings $\bar{\textbf{E}}^{(0)}$, the number of graph layer $L$, the number samples for training $S$, weight $\lambda$ for regularization, the number of epochs $E$}
	\KwOut{trained model parameters $\mathbf{\Theta}$}
	hyperparameter Initializations $\mathbf{\Theta}$\\
	\For{$e=1$ to $E$}{
	    Seed node sampling $\mathbb{U}$, $\mathbb{V}$\\
    	sub-graph generation $(\hat{\mathbb{U}}, \hat{\mathbb{V}}, \hat{\textbf{X}})$ using the seeds according to Algorithm~\ref{alg:sample}\\
    	Get $\textbf{E}^{(0)}$ from $\bar{\textbf{E}}^{(0)}$ for $u_i$ in $\mathbb{U}$ and $v_j$ in $\mathbb{V}$\\
    	\For{$l=1$ to $L$}{
    	    \For{each $u_i$ in $\hat{\mathbb{U}}$, $v_j$ in $\hat{\mathbb{V}}$ and $k=1$ to $K$}{
    	        Type-specific behavior message generation   $\textbf{H}^k$\\
    	        Message refinement $\hat{\textbf{H}}^k$\\
    	        Embedding aggregation $\textbf{E}^{(l)}$\\
        	}
    	}
    	
    	$\mathcal{L}=\lambda\|\mathbf{\Theta}_\text{F}^2\|$\\
    	\For{each $u_i$ in $\hat{\mathbb{U}}$}{
    	   Positive and negative items are sampled from $\hat{\mathbb{V}}$\\
    	    \For{each $v_{p_s}$ and ${v_{n_s}}$} {
    	       Interaction probability inference $\text{Pr}_{i,j}$.\\
    	       $\mathcal{L}+=\max(1-\text{Pr}_{i,p_s}+\text{Pr}_{i,n_s})$
    	    }
    	}
        	Model training with the optimized objective. $\mathcal{L}$\\
    }
    return $\mathbf{\Theta}$
	
\end{algorithm}

To perform the model inference, we optimize our \model\ framework with the pair-wise loss, which has been widely adopted in the Top-N recommendation scenarios~\cite{nikolakopoulos2019recwalk}.
During the training phase, $S$ positive instances and $S$ negative instances will be sampled from the observed interacted item set and non-interacted item set, respectively. Our model parameters are inferred by minimizing the defined loss function:
\begin{align}
    \mathcal{L}=\sum_{i=1}^I\sum_{s=1}^S\max(1-\text{Pr}_{i,p_s}+\text{Pr}_{i,n_s}, 0)+\lambda\|\mathbf{\Theta}\|_{\text{F}}^2
\end{align}
where the first term is the pair-wise loss, and the second term is the regularization term with hyper-parameter $\lambda$ as weight. The set of model parameters is denoted by $\mathbf{\Theta}$. We summarize the procedure of model training with the sub-graph sampling algorithm in Algorithm~\ref{alg:train}.

\subsection{Model Complexity Analysis}
\label{sec:complexity}
\noindent \textbf{Time Complexity.} The running time of our model can be divided into two parts: the time for sub-graph sampling, and the cost for model training and inference. As described in Algorithm~\ref{alg:sample}, the major cost of the former process is $O(D\times N\times (I + J))$ for updating the sampling probability $P_u$ and $P_v$, where $D$ is the number of sampling steps and $N$ is the number of sampled nodes per step. In the model running phase, \model\ takes $O(L\times |\textbf{X}|\times d)$ ($|\textbf{X}|$ denotes the number of non-zero elements in $\textbf{X}$) to encode the type-specific message, in which $O(L\times (I+J)\times K\times d^2)$ is needed by the attention module. The complexity of the type-wise inter-dependency modeling and the aggregation layer is analogously $O(L\times (I+J)\times K\times d^2)$, in which the primary contributor is the matrix-multiplications. The complexity of the cross-order mutual relation learning is $O(|\textbf{X}|\times L^2\times d^2)$ which comes from the order-wise representation fusion, and this term dominates the complexity of the model running process. Empirically, by sharing the sampled sub-graph among training/testing steps, the sub-graph sampling costs much less time compared to the entire computational cost.

\noindent \textbf{Space Complexity.} Due to sampling larger sub-graphs for computing efficiency and data integrity, \model\ takes more memory than some GNN model in sub-graph sampling. But the memory cost is fully acceptable for common devices. For the model memory cost, the space complexity of \model\ is $O(L\times (I+J)\times K\times d)$, which is mainly for the intermediate hidden states, the same as a common graph neural networks (\eg~GCN and GraphSAGE) for modeling multi-behavior data.

\section{Evaluation}
\label{sec:eval}

We evaluate \model\ to answer the research questions as:

\begin{itemize}[leftmargin=*]

\item \textbf{RQ1}: How does our \emph{\model} perform compared with various recommendation baselines on different datasets? \\\vspace{-0.1in}

\item \textbf{RQ2}: How does each model design (\eg, multi-channel behavior embedding layer and cross-layer mutual relation learning module) affect the model performance? \\\vspace{-0.1in}

\item \textbf{RQ3}: What is the impact of incorporating different types of behaviour context in our graph neural multi-behavior recommender system? \\\vspace{-0.1in}

\item \textbf{RQ4}: How do different interaction sparsity degrees influence the recommendation performance? \\\vspace{-0.1in}

\item \textbf{RQ5}: How do the key hyperparameters impact the performance of \emph{\model} neural architecture? \\\vspace{-0.1in}

\item \textbf{RQ6}: How is the model efficiency of \emph{\model} when competing with various types of recommendation techniques? \\\vspace{-0.1in}

\item \textbf{RQ7}: How does the user multi-behavior dependency study benefit the interpretation ability for recommendation? \\\vspace{-0.1in}

\item \textbf{RQ8}: What is the effect of the graph sampling algorithm on the model performance of \emph{\model}?

\end{itemize}


\begin{table}[t]
    \caption{Statistics of our evaluation datasets.}
    \vspace{-0.1in}
    \label{tab:data}
    \centering
    \scriptsize
	\setlength{\tabcolsep}{0.6mm}
    \begin{tabular}{ccccc}
        \toprule
        Dataset&User \#&Item \#&Interaction \#&Interactive Behavior Type\\
        \midrule
        Tmall & 147894 & 99037 & 7658926 & \{Page View, Favorite, Cart, Purchase\}\\
        BeiBei & 21716 & 7977 & 3338068 &\{Page View, Cart, Purchase\}\\
        IJCAI & 423423 & 874328 & 36203512 &\{Page View, Favorite, Cart, Purchase\}\\
        \hline
    \end{tabular}
\vspace{-0.15in}
\end{table}

\begin{table*}[t]
\caption{Performance comparison on Beibei, Tmall and IJCAI data, in terms of \textit{HR@$N$} and \textit{NDCG@$N$} ($N=10$).}
\centering
\scriptsize
\setlength{\tabcolsep}{1mm}
\begin{tabular}{|c|c|c|c|c|c|c|c|c|c|c|c|c|c|c|c|c|c|}
\hline
Data & Metric & BiasMF & DMF & NCF & AutoRec & CDAE & NADE & CF-UIcA & ST-GCN & NGCF & NMTR & DIPN & NGCF+M & MBGCN & MATN & GNMR & \emph{\model}\\
\hline
\multirow{4}{*}{Beibei}
&HR & 0.588 & 0.597 & 0.595 & 0.607 & 0.608 & 0.608 & 0.610 & 0.609 & 0.611 & 0.613 & 0.631 & 0.634 & 0.642 & 0.626 & 0.631 & \textbf{0.670}\\
\cline{2-18}
&Imprv & 13.95\% & 12.23\% & 12.61\% & 10.38\% & 10.20\% & 10.20\% & 9.84\% & 10.02\% & 9.66\% & 9.30\% & 6.18\% & 5.68\% & 4.36\% & 7.03\% & 6.18\% & --\\
\cline{2-18}
&NDCG & 0.333 & 0.336 & 0.332 & 0.341 & 0.341 & 0.343 & 0.346 & 0.343 & 0.375 & 0.349 & 0.384 & 0.372 & 0.376 & 0.385 & 0.380 & \textbf{0.402}\\
\cline{2-18}
&Imprv & 20.72\% & 19.64\% & 21.08\% & 17.89\% & 17.89\% & 17.20\% & 16.18\% & 17.20\% & 7.20\% & 15.19\% & 4.69\% & 8.06\% & 6.91\% & 4.42\% & 5.79\% & --\\
\hline
\multirow{4}{*}{Tmall}
&HR & 0.262 & 0.305 & 0.319 & 0.313 & 0.329 & 0.317 & 0.332 & 0.347 & 0.302 & 0.332 & 0.317 & 0.374 & 0.369 & 0.354 & 0.424 & \textbf{0.444} \\
\cline{2-18}
&Imprv & 69.47\% & 45.57\% & 39.18\% & 41.85\% & 34.95\% & 40.06\% & 33.73\% & 27.95\% & 47.02\% & 33.73\% & 40.06\% & 18.72\% & 20.33\% & 25.42\% & 4.72\% & --\\
\cline{2-18}
&NDCG & 0.153 & 0.189 & 0.191 & 0.190 & 0.196 & 0.191 & 0.198 & 0.206 & 0.185 & 0.179 & 0.178 & 0.221 & 0.222 & 0.209 & 0.249 & \textbf{0.262} \\
\cline{2-18}
&Imprv & 71.24\% & 38.62\% & 37.17\% & 37.89\% & 33.67\% & 37.17\% & 32.32\% & 27.18\% & 41.62\% & 46.37\% & 47.19\% & 18.55\% & 18.02\% & 25.36\% & 5.22\% & --\\
\hline
\multirow{4}{*}{IJCAI}
&HR & 0.285 & 0.392 & 0.449 & 0.448 & 0.455 & 0.469 & 0.429 & 0.452 & 0.461 & 0.481 & 0.475 & 0.481 & 0.463 & 0.489 & 0.519 & \textbf{0.554}\\
\cline{2-18}
&Imprv & 94.39\% & 41.33\% & 23.39\% & 23.66\% & 21.76\% & 18.12\% & 29.14\% & 22.57\% & 20.17\% & 15.18\% & 16.63\% & 15.18\% & 19.65\% & 13.29\% & 6.74\% & --\\
\cline{2-18}
&NDCG & 0.185 & 0.250 & 0.284 & 0.287 & 0.288 & 0.304 & 0.260 & 0.285 & 0.292 & 0.304 & 0.296 & 0.307 & 0.277 & 0.309 & 0.312 & \textbf{0.338}\\
\cline{2-18}
&Imprv & 82.70\% & 35.20\% & 19.01\% & 17.77\% & 17.36\% & 11.18\% & 30.00\% & 18.60\% & 15.75\% & 11.18\% & 14.19\% & 10.10\% & 22.02\% & 9.39\% & 8.33\% & --\\
\hline
\end{tabular}
\vspace{-0.05in}
\label{tab:target_behavior}
\end{table*}

\subsection{Data Description}
Our evaluations are performed on three real-world datasets: Tmall, BeiBei and IJCAI-Competition. We summarize the detailed statistical information of those datasets in Table~\ref{tab:data}.

\begin{itemize}[leftmargin=*]

\item \textbf{Tmall}. This is a public recommendation dataset from the Tmall e-commerce platform by including four types of user behaviors: click, add-to-cart, tag-as-favorite and purchase. This data contains 47,894 users and 99,037 items.\\\vspace{-0.12in}

\item \textbf{BeiBei}. This is another e-commerce dataset for item recommendation from one of the largest infant product retail site in China. There are 21,716 users and 7,977 items in this dataset with three types of user-item interactions, namely, click, add-to-cart and purchase.\\\vspace{-0.12in}

\item \textbf{IJCAI-Competition}. This data comes from the released repository of IJCAI competition to provide researchers with user online behavior modeling. It involves four types of interaction behavior between user and item, \ie, click, add-to-cart, tag-as-favorite and purchase. 423,423 users and 874,328 items are included in this data source.

\end{itemize}

To be consistent with the settings in~\cite{matnsigir20,jin2020multi}, the target predicted behaviors in our recommendation scenario are user purchases and other types of behaviors (\eg, click, add-to-cart) are regarded as the auxiliary behaviours.

\subsection{Evaluation Metrics}
In our experiments, the models are evaluated on the top-$N$ item recommendation task with the metrics of Hit Ratio (HR)@$N$ and NDCG@$N$. In our evaluation protocol, we use the leave one item out strategy~\cite{zhao2020revisiting} to consider the last interaction with the target behavior of each user as the testing set. In particular, following the similar settings in~\cite{kang2018self,sun2019bert4rec}, for individual user, we sample 99 items as negative instances from the set of all non-interacted items. Items in the test set are regarded as the positive instances. 

\subsection{Baseline Models}
To demonstrate the effectiveness of our \emph{\model} framework, we compare our method with the following state-of-the-art methods, which involves different categories:

\noindent \textbf{Conventional Matrix Factorization Method}:
\begin{itemize}[leftmargin=*]
\item \textbf{BiasMF}~\cite{koren2009matrix}: this model attempts to incorporate user and item bias information into the matrix factorization, so as to learn latent embeddings of users/items.
\end{itemize}

\noindent \textbf{Neural Collaborative Filtering}:
\begin{itemize}[leftmargin=*]

\item \textbf{NCF}~\cite{he2017neuralncf}: it augments the embedding paradigm in collaborative filtering with the multilayer perceptron to enable the non-linear feature interactions.

\item \textbf{DMF}~\cite{xue2017deep}: this is another neural collaborative filtering technique, to learn a common low dimensional space for users and items with non-linear transformations.

\end{itemize}

\noindent \textbf{Autoencoder-based Recommendation Models}:
\begin{itemize}[leftmargin=*]

\item \textbf{AutoRec}~\cite{sedhain2015autorec}: this recommendation model stacks multiple autoencoder layers to project user-item interaction inputs into the latent representations for data reconstruction.


\item \textbf{CDAE}~\cite{wu2016collaborative}: It is a model-based CF recommender with the denoising auto-encoder technique to learn user correlations.

\end{itemize}

\noindent \textbf{Neural Auto-regressive Recommender Systems}:

\begin{itemize}[leftmargin=*]

\item \textbf{NADE}~\cite{zheng2016neural}: it designs a neural autoregressive architecture for recommendation task with the parameter sharing between different ratings.

\item \textbf{CF-UIcA}~\cite{du2018collaborative}: this is an user-item co-autoregressive framework with a new stochastic learning strategy to encode correlations between users and items.

\end{itemize}

\noindent \textbf{Graph Neural Network-based Recommendation Methods}:
\begin{itemize}[leftmargin=*]

\item \textbf{ST-GCN}~\cite{zhang2019star}: this graph-based method is built over an encoder-decoder framework to perform the convolution-based embedding propagation between user and item nodes.

\item \textbf{NGCF}~\cite{wang2019neural}: it is a state-of-the-art GNN-based colloborative filtering model which exploits the high-order user-item interaction structures.

\end{itemize}

\noindent \textbf{Multi-Behavior Recommender Systems}:
\begin{itemize}[leftmargin=*]


\item \textbf{NMTR}~\cite{gao2019neural}: This method relies on the defined cascaded behavior relationships for encoding the multi-behavior semantics with a multi-task learning scheme.


\item \textbf{DIPN}~\cite{guo2019buying}: this deep intent prediction network aims to integrate the browsing and buying preferences of users with a new type touch-interactive behavior patterns.

\item \textbf{NGCF+M}~\cite{wang2019neural}: we generate a new multi-behavior recommendation variant of NGCF by injecting the multi-behavior context into the message passing scheme.

\item \textbf{MATN}~\cite{matnsigir20}: this recommendation model considers the influences among different types of interactions with attentive weights for pattern aggregation.

\item \textbf{GNMR}~\cite{icdegnmr}: this is the previous version of our \emph{\model} which captures the pairwise dependencies between different types of behaviors with the integration of the multi-channel behavior representation layer and self-attention network for relation aggregation. However, it ignores the layer-wise embedding dependency during the representation integration.

\item \textbf{MBGCN}~\cite{jin2020multi}: this multi-behavior recommender system leverages the graph convolutional network to capture the multi-behaviour patterns over the interaction graph.

\end{itemize}


\subsection{\bf Parameter Settings}
Our \emph{\model} model is implemented with TensorFlow. The parameter inference is conducted with the Adam optimizer and the training phase is performed with the learning rate of $1e^{-3}$ and batch size of 32. For the model hyperparameters, the dimensionality of hidden state $d$ is set as 16 in our representation space. The number of channels for behavior embedding layer is set as 8. We use 2 attention-based representation heads in our behavior inter-dependency modeling component. To alleviate the overfitting issue, the regularization strategy with the weight decay parameter sampled from \{0.05, 0.01, 0.005, 0.001\}. 


\begin{table}[t]
	\caption{Recommendation accuracy with different Top-\textit{N} values in terms of \textit{HR@N} and \textit{NDCG@N} on BeiBei dataset.}
	\centering
    \scriptsize
	\setlength{\tabcolsep}{1mm}
	\begin{tabular}{|c|c|c|c|c|c|c|c|c|}
		\hline
		\multirow{2}{*}{Model}&\multicolumn{2}{c|}{@5}&\multicolumn{2}{c|}{@10}&\multicolumn{2}{c|}{@20}&\multicolumn{2}{c|}{@50} \\
		\cline{2-9}
		&HR&NDCG&HR&NDCG&HR&NDCG&HR&NDCG\\
		\hline
		\hline
		BiasMF & 0.453 & 0.287 & 0.588 & 0.333 & 0.678 & 0.357 & 0.807 & 0.379\\
		\hline
        NCF & 0.447 & 0.283 & 0.601 & 0.336 & 0.698 & 0.359 & 0.819 & 0.383\\
        \hline
        NGCF+M & 0.496 & 0.337 & 0.634 & 0.372 & 0.743 & 0.381 & 0.872 & 0.407\\
        \hline
        MBGCN & 0.498 & 0.337 & 0.642 & 0.376 & 0.740 & 0.398 & 0.902 & 0.429\\
        \hline
        AutoRec & 0.456 & 0.291 & 0.607 & 0.341 & 0.707 & 0.366 & 0.826 & 0.391\\
        \hline
        MATN & 0.467 & 0.330 & 0.626 & 0.385 & 0.667 & 0.342 & 0.833 & 0.396\\
        \hline
        \emph{\model} & \textbf{0.527} & \textbf{0.359} & \textbf{0.670} & \textbf{0.402} & \textbf{0.788} & \textbf{0.433} & \textbf{0.927} & \textbf{0.461}\\
		\hline
	\end{tabular}
	\label{tab:vary_k}
\end{table}

\begin{table}[t]
	\caption{Performance comparison with different number of negative samples in terms of \textit{HR@10} and \textit{NDCG@10}}
	\centering
    \scriptsize
	\setlength{\tabcolsep}{0.6mm}
	\begin{tabular}{|c|c|c|c|c|c|c|c|c|c|c|}
		\hline
		\# samples &\multicolumn{2}{c|}{400}&\multicolumn{2}{c|}{800}&\multicolumn{2}{c|}{1600}&\multicolumn{2}{c|}{3200} & \multicolumn{2}{c|}{6400}\\
		\hline
		Model&HR&NDCG&HR&NDCG&HR&NDCG&HR&NDCG & HR & NDCG\\
		\hline
		\hline
		BiasMF & 0.285 & 0.140 & 0.149 & 0.078 & 0.091 & 0.049 & 0.055 & 0.032 & 0.036 & 0.022\\
		\hline
		NCF & 0.295 & 0.144 & 0.155 & 0.080 & 0.090 & 0.047 & 0.053 & 0.029 & 0.037 & 0.019\\
		\hline
        AutoRec & 0.245 & 0.122 & 0.134 & 0.074 & 0.0761 & 0.047 & 0.0500 & 0.033 & 0.036 & 0.023\\
        \hline
        ST-GCN & 0.311 & 0.158 & 0.184 & 0.093 & 0.104 & 0.053 & 0.058 & 0.031 & 0.036 & 0.019\\
        \hline
        MBGCN & 0.353 & 0.181 & 0.218 & 0.099 & 0.122 & 0.063 & 0.073 & 0.035 & 0.039 & 0.018\\
        \hline
        MATN & 0.339 & 0.165 & 0.192 & 0.093 & 0.113 & 0.056 & 0.058 & 0.031 & 0.037 & 0.020\\
        \hline
        GNMR & 0.345 & 0.176 & 0.218 & 0.100 & 0.103 & 0.055 & 0.063 & 0.036 & 0.040 & 0.024\\
        \hline
        \emph{\model} & \textbf{0.361} & \textbf{0.1848} & \textbf{0.219} & \textbf{0.109} &\textbf{0.124} & \textbf{0.064} & \textbf{0.075} & \textbf{0.040} & \textbf{0.044} & \textbf{0.025}\\
		\hline
	\end{tabular}
	\vspace{-0.1in}
	\label{tab:neg_samp}
\end{table}

\subsection{Performance Comparison (RQ1)}
The evaluation results (measured by HR@10 and NDCG@10) of all compared methods on three datasets are shown in Table~\ref{tab:target_behavior}. In all cases, we could observe that \emph{\model} consistently outperforms baseline methods from various research lines by a significant margin. We attribute such performance improvement to the joint learning of multi-behavior inter-dependencies as well as the cross-layer collaborative signals under graph neural network. For example, \emph{\model} makes over 34\% and 33\% relatively improvement with respect to HR@10 and NDCG@10 respectively, as compared to autoencoder-based recommendation models (\ie, AutoRec \& CDAE) on Tmall data. Additionally, for the results in terms of HR@10 on IJCAI-Competition data, the constant gain achieved by the developed \emph{\model} is around 20-22\% over graph neural network-based CF models (ST-GCN and NGCF), and 18-29\% over neural auto-regressive recommendation methods (NADE and CF-UIcA).

The proposed \emph{\model} also outperforms all other baseline methods with the modeling of multi-behavior data with respect to all metrics. Results show that our \emph{\model} allows the graph neural architecture to capture the multi-behavior interaction patterns, and successfully distinguish the layer-wise representations. While MBGCN and NGCF+M are built over the graph neural network to model behavior correlations, they fall short in encoding the latent type-specific characteristics and cross-type behavior inter-dependencies simultaneously. The performance of NMTR and MATN are limited to the failure for considering the high-order collaborative effects over the multi-behavior interaction graph. Furthermore, our new version model \emph{\model} always achieves better recommendation accuracy than the simplified version GNMR, which also confirms the effectiveness of our designed component for cross-layer mutual relation modeling. We further evaluate the performance of our \emph{\model} and several representative baselines with different top-$N$ positions. The results are reported in Table~\ref{tab:vary_k}. The best performance is achieved by our framework under different settings.
To further evaluate the performance of our \model\ framework, we make the performance comparison by varying the number of sample negative instances. The evaluation results are shown in Table~\ref{tab:neg_samp}. We can observe that our \model\ method consistently outperforms other alternative methods under different settings of negative samples in the range of \{400, 800, 1600, 3200, 6400\}. This observation validates the superiority of our \model\ in advancing the recommendation performance with the effective modeling of high-order heterogeneous collaborative relationships.


\subsection{Ablation Study (RQ2)}
In this section, we would like to answer the question that if the designed individual component could help improve the recommendation accuracy. Specifically, we generate four types of model variants of our \emph{\model} corresponding to different aspects of our model design:

\begin{itemize}[leftmargin=*]

\item \textbf{Impact of Multi-Channel Behavior Embedding}. To evaluate the effect of our multi-channel behavior embedding layer, we compare the proposed method with the variant (w/o-MCE). This variant discards the behavior semantic modeling with multi-channel representation spaces. As shown in Table~\ref{tab:module_ablation} about the evaluation on three datasets, we can observe that the results of \emph{\model} are better than that of the variant (w/o-MCE). It demonstrates that the encoding of type-specific behavior characteristic could facilitate the multi-behavior dependency modeling. \\\vspace{-0.1in}

\item \textbf{Impact of Behavior Inter-dependency Modeling}.
To investigate the rationality of our behavior inter-dependency modeling, our \emph{\model} is compared with another model implementation (w/o-BIM) by removing the multi-behavior attention network. From the results in Table~\ref{tab:module_ablation}, \emph{\model} outperforms w/o-BIM in all cases, which benefits from the user/item representation enhanced by the exploration of pairwise behavior relational structures. \\\vspace{-0.1in}

\item \textbf{Impact of Behavior Pattern Fusion}. 
We generate another simplified implementation of our recommendation architecture: (w/o-BFu) that does not consider the aggregation layer for pattern aggregation across various types of behavior representations. Instead, the type-aware behavior representations are directly combined through the element-wise mean pooling. As expected, \emph{\model} achieves better recommendation accuracy as compared to the variant (w/o-BFu). It verifies the necessity of our embedding fusion scheme during our multi-behavior dependency modeling. \\\vspace{-0.1in}

\item \textbf{Impact of High-order Mutual Relation Learning}.
To evaluate the effect of augmenting the graph neural model by capturing the cross-layer collaborative relations, we generate another variant (w/o-HMR) by only generating the output from the highest graph order after the information propagation process. From the evaluation results, we can observe the efficacy of the designed mutual relation encoder in learning the contributions of order-specific embeddings for the final prediction result.

\end{itemize}

\begin{table}[t]
    \caption{Ablation study on key components of \model.}
    \centering
    \footnotesize
    \setlength{\tabcolsep}{1mm}
    \begin{tabular}{c|cc|cc|cc}
        \hline
        Data & \multicolumn{2}{c|}{Beibei Data} & \multicolumn{2}{c|}{Tmall Data} & \multicolumn{2}{c}{IJCAI Data}\\
        \hline
        Metrics & HR & NDCG & HR & NDCG & HR & NDCG\\
        \hline
        \hline
        w/o-MCE & 0.6549 & 0.3876 & 0.4399 & 0.2580 & 0.5420 & 0.3289\\ 
        w/o-BIM & 0.6696 & 0.4000 & 0.4391 & 0.2554 & 0.5358 & 0.3228\\ 
        w/o-BFu & 0.6572 & 0.3907 & 0.4238 & 0.2487 & 0.5494 & 0.3321\\ 
        w/o-HMR & 0.6169 & 0.3470 & 0.3856 & 0.2240 & 0.3445 & 0.1760\\
        \hline
        \emph{\model} & \textbf{0.6701} & \textbf{0.4021} & \textbf{0.4435} & \textbf{0.2624} & \textbf{0.5535} & \textbf{0.3376}\\
        \hline
    \end{tabular}
    \label{tab:module_ablation}
\end{table}

\subsection{Analysis on Individual Behavior Context (RQ3)}
\begin{figure}[t]
	\centering
	\subfigure[][Beibei-HR]{
		\centering
		\includegraphics[width=0.28\columnwidth]{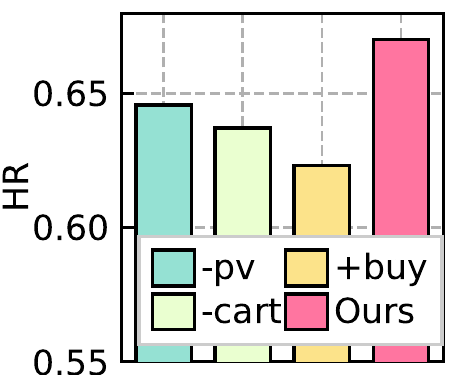}
		\label{fig:ab_beibei_hr}
	}
	\subfigure[][IJCAI-HR]{
		\centering
		\includegraphics[width=0.28\columnwidth]{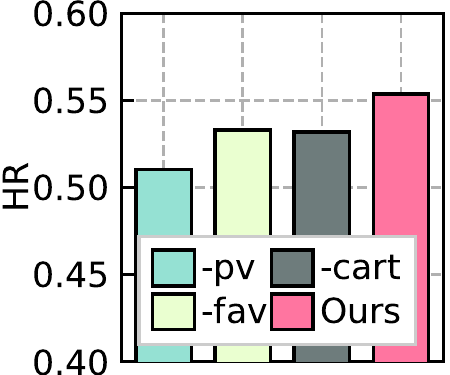}
		\label{fig:ab_ijcai_HR}
	}
	\subfigure[][Tmall-HR]{
		\centering
		\includegraphics[width=0.28\columnwidth]{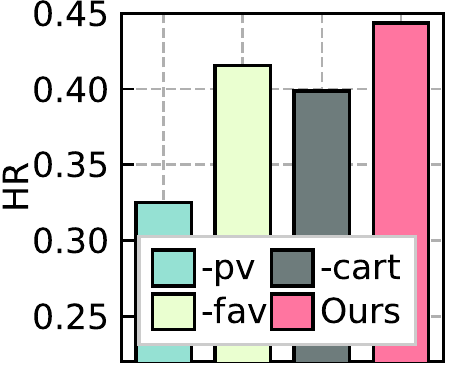}
		\label{fig:ab_tmall_HR}
	}
	\subfigure[][Beibei-NDCG]{
		\centering
		\includegraphics[width=0.28\columnwidth]{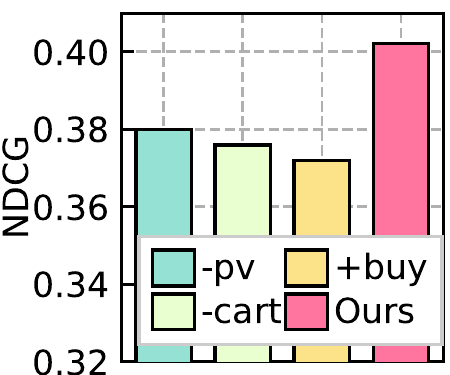}
		\label{fig:ab_beibei_NDCG}
	}
	\subfigure[][IJCAI-NDCG]{
		\centering
		\includegraphics[width=0.28\columnwidth]{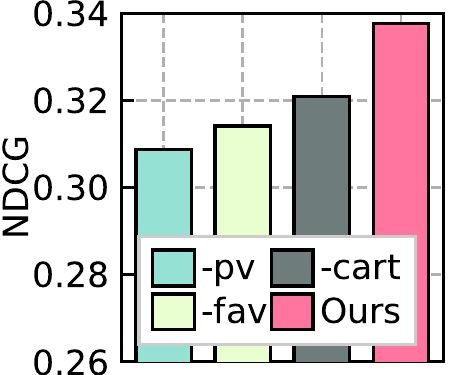}
		\label{fig:ab_ijcai_NDCG}
	}
	\subfigure[][Tmall-NDCG]{
		\centering
		\includegraphics[width=0.28\columnwidth]{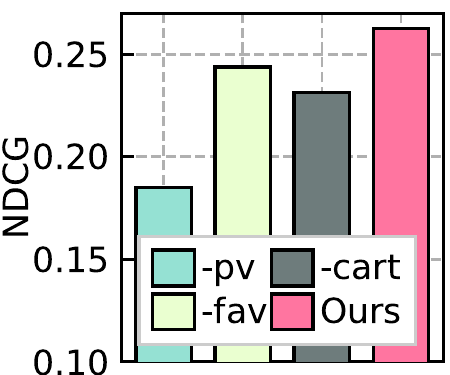}
		\label{fig:ab_tmall_NDCG}
	}
	\vspace{-0.1in}
	\caption{Impact study of diverse behavior types. There are three types of behaviors for BeiBei data, and four types of behaviors for Tmall, IJCAI data.}
	\label{fig:beh_ablation}
	\vspace{-0.1in}
\end{figure}

This section conducts ablation studies on the influence of type-specific behavior context for the recommendation performance. The compared model variants are generated with the rubric as: First, ``+'' behavior type means that merely considering the target behaviors into the system to make predictions (\ie, +buy). Second, ``-'' behavior type indicates the removing of this certain type of user behaviors (\eg, -pv, -cart) from the recommendation architecture. For instance, -pv indicates that we do not include the page view behaviors into the interaction inter-dependency modeling. We present the evaluation results in terms of NDCG@N and HR@N when $N=10$ on three real-world datasets in Figure~\ref{fig:beh_ablation}. As shown in Table~\ref{tab:data}, the number of behavior types is 3 (\ie, page view, add-to-cart, buy) on BeiBei data and 4 (\ie, page view, add-to-cart, tag-as-favorite, buy) on Tamll, IJCAI-Competition data. From the results, we can observe that each type of interaction behavior individually contributes to improve the user preference learning, and integrate multi-behavior behavior patterns for performance improvement.

\subsection{Performance Under Different Sparsity (RQ4)}
\label{sec:sparsity}
\begin{figure}[t]
	\centering
	\subfigure[][Tmall HR@10]{
		\centering
		\includegraphics[width=0.45\columnwidth]{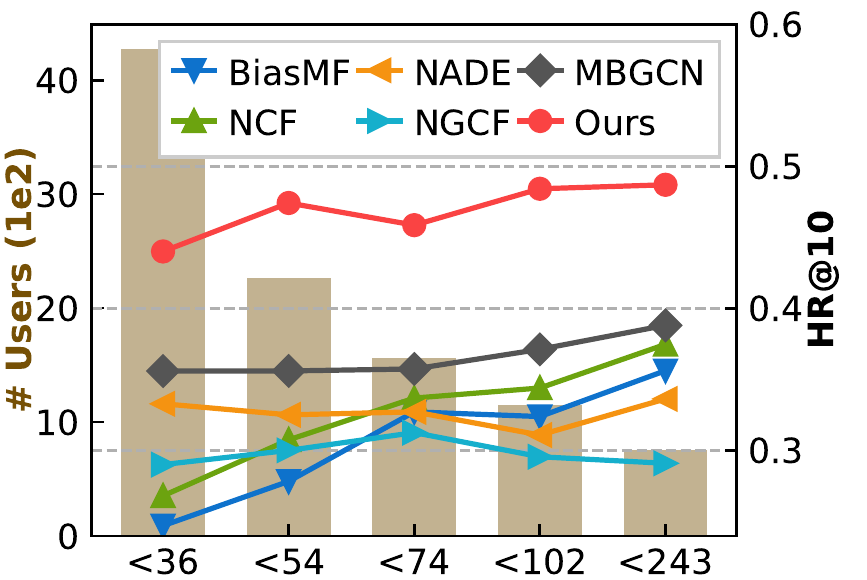}
		\label{fig:sp_tmall_hr}
	}
	\subfigure[][Tmall NDCG@10]{
		\centering
		\includegraphics[width=0.45\columnwidth]{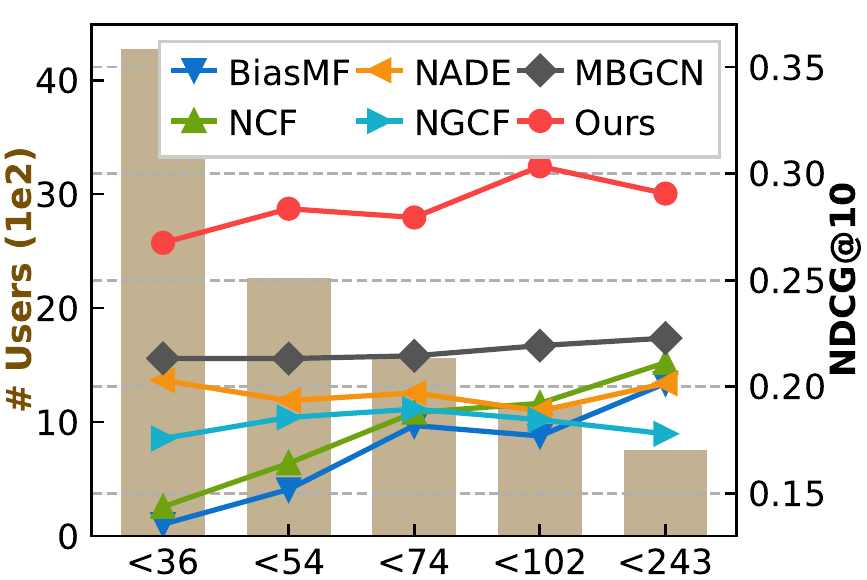}
		\label{fig:sp_tmall_ndcg}
	}
	\caption{Performance comparison of \model\ and baseline methods \wrt\ different data sparsity levels on Tmall data.}
	\label{fig:sparsity}
\end{figure}

In our experiments, we also evaluate the recommendation performance of different models under different interaction sparsity. Following the similar settings in~\cite{wang2019neural,wu2020diffnet}, we first partition users into five groups based on the number of interactions. For example, ``\textless36'' and ``\textless52'' indicate that users belong to this group have the number of interactions ranging from 1 to 35, and 36 to 51, respectively. We keep the same number of users in each group and select the corresponding ranges as shown in x-axis of Figure~\ref{fig:sparsity}. The total number of users contained in each group and the recommendation accuracy with respect to HR (Figure~\ref{fig:sparsity} (a)) and NDCG (Figure~\ref{fig:sparsity} (b)) are shown in the left side and right side of y-axis in Figure~\ref{fig:sparsity}. From evaluation results, we can notice the superiority of our \emph{\model} with different sparsity levels. It suggests that the incorporation of multi-typed behaviour patterns into the user preference learning could reach performance improvement as compared with other baselines. In addition, we can observe that the overall performance of all compared methods share similar increase trend as users have more interactions. This may indicate that more user behavior data may help characterize user preference with more accurate latent representations.

\begin{table}[t]
    \caption{Effect of embedding propagation layers.}
    \centering
    \footnotesize
    \begin{tabular}{c|cc|cc|cc}
        \hline
        Data & \multicolumn{2}{c|}{BeiBei Data} & \multicolumn{2}{c|}{Tmall Data} & \multicolumn{2}{c}{IJCAI Data}\\
        \hline
        Metrics & HR & NDCG & HR & NDCG & HR & NDCG\\
        \hline
        \hline
        \emph{\model}-1 & 0.662 & 0.394 & 0.383 & 0.226 & 0.533 & 0.313 \\
        \emph{\model}-2 & \textbf{0.670} & \textbf{0.402} & \textbf{0.444} & \textbf{0.262} & \textbf{0.554} & \textbf{0.338} \\
        \emph{\model}-3 & 0.664 & 0.398 & 0.408 & 0.237 & 0.543 & 0.332 \\
        \hline
    \end{tabular}
    \vspace{-0.05in}
    \label{tab:graph_layers}
\end{table}

\begin{table}[t]
    \caption{Effect of behavior embedding channels.}
    \vspace{-0.1in}
    \centering
    \footnotesize
    \begin{tabular}{c|cc|cc|cc}
        \hline
        Data & \multicolumn{2}{c|}{BeiBei Data} & \multicolumn{2}{c|}{Tmall Data} & \multicolumn{2}{c}{IJCAI Data}\\
        \hline
        Metrics & HR & NDCG & HR & NDCG & HR & NDCG\\
        \hline
        \hline
        \emph{\model}-2 & 0.655 & 0.390 & 0.438 & 0.260 & 0.518 & 0.308 \\
        \emph{\model}-4 & 0.656 & 0.401 & 0.443 & 0.260 & 0.553 & 0.328 \\
        \emph{\model}-8 & \textbf{0.670} & \textbf{0.402} & \textbf{0.444} & \textbf{0.262} & \textbf{0.554} & \textbf{0.338} \\
        \emph{\model}-16 & 0.646 & 0.387 & 0.419 & 0.241 & 0.558 & 0.336 \\
        \hline
    \end{tabular}
    \label{tab:memory_units}
\end{table}

\subsection{Analysis on Hyperparameters (RQ5)}
\label{sec:hyper_study}
We study the impact of different hyperparameter settings on the model performance in our joint learning framework.\\\vspace{-0.1in}

\begin{figure}[t]
\vspace{-0.05in}
    \centering
    \begin{adjustbox}{max width=1.0\linewidth}
    \begin{tikzpicture}
\begin{axis}[
    xlabel={Hidden State Dimensionality $d$},
    ylabel={Hit Rate@10},
    xmin=1,xmax=33,
    ymin=0.62,ymax=0.68,
    legend columns=1,
    legend cell align=right,
    grid=both,
    every axis plot/.append style={ultra thick},
    every tick label/.append style={scale=1.3},
    label style={scale=1.8},
    legend style={
        nodes={scale=1.5, transform shape},
        legend image post style={scale=1.5},
        },
    legend style={at={(1,0)},anchor=south east},
    every axis plot post/.append style={
        every mark/.append style={scale=2}
    }
]
\addplot[color={rgb:red,0;green,157;blue,178}, mark=square, mark options={solid}]
table[x=para, y=beibei_hr] {latFactor.txt};
\legend{\large Beibei};
\end{axis}
\end{tikzpicture}

\begin{tikzpicture}
\begin{axis}[
    xlabel={Hidden State Dimensionality $d$},
    ylabel={Hit Rate@10},
    xmin=1,xmax=33,
    ymin=0.31,ymax=0.46,
    legend columns=1,
    legend cell align=right,
    grid=both,
    every axis plot/.append style={ultra thick},
    every tick label/.append style={scale=1.3},
    label style={scale=1.8},
    legend style={
        nodes={scale=1.5, transform shape},
        legend image post style={scale=1.5},
        },
    legend style={at={(1,0)},anchor=south east},
    every axis plot post/.append style={
        every mark/.append style={scale=2}
    }
]
\addplot[color={rgb:red,0;green,157;blue,178}, mark=square, mark options={solid}]
table[x=para, y=tmall_hr] {latFactor.txt};
\legend{\large Tmall};
\end{axis}
\end{tikzpicture}

\begin{tikzpicture}
\begin{axis}[
    xlabel={Hidden State Dimensionality $d$},
    ylabel={Hit Rate@10},
    xmin=1,xmax=33,
    ymin=0.45,ymax=0.56,
    legend columns=1,
    legend cell align=right,
    grid=both,
    every axis plot/.append style={ultra thick},
    every tick label/.append style={scale=1.3},
    label style={scale=1.8},
    legend style={
        nodes={scale=1.5, transform shape},
        legend image post style={scale=1.5},
        },
    legend style={at={(1,0)},anchor=south east},
    every axis plot post/.append style={
        every mark/.append style={scale=2}
    }
]
\addplot[color={rgb:red,0;green,157;blue,178}, mark=square, mark options={solid}]
table[x=para, y=ijcai_hr] {latFactor.txt};
\legend{\large IJCAI};
\end{axis}
\end{tikzpicture}
    \end{adjustbox}
    \begin{adjustbox}{max width=1.0\linewidth}
    \begin{tikzpicture}
\begin{axis}[
    xlabel={Hidden State Dimensionality $d$},
    ylabel={NDCG@10},
    xmin=1,xmax=33,
    ymin=0.35,ymax=0.41,
    legend columns=1,
    legend cell align=right,
    grid=both,
    every axis plot/.append style={ultra thick},
    every tick label/.append style={scale=1.3},
    label style={scale=1.8},
    legend style={
        nodes={scale=1.5, transform shape},
        legend image post style={scale=1.5},
        },
    legend style={at={(1,0)},anchor=south east},
    every axis plot post/.append style={
        every mark/.append style={scale=2}
    }
]
\addplot[color={rgb:red,0;green,157;blue,178}, mark=square, mark options={solid}]
table[x=para, y=beibei_ndcg] {latFactor.txt};
\legend{\large Beibei};
\end{axis}
\end{tikzpicture}

\begin{tikzpicture}
\begin{axis}[
    xlabel={Hidden State Dimensionality $d$},
    ylabel={NDCG@10},
    xmin=1,xmax=33,
    ymin=0.17,ymax=0.27,
    legend columns=1,
    legend cell align=right,
    grid=both,
    every axis plot/.append style={ultra thick},
    every tick label/.append style={scale=1.3},
    label style={scale=1.8},
    legend style={
        nodes={scale=1.5, transform shape},
        legend image post style={scale=1.5},
        },
    legend style={at={(1,0)},anchor=south east},
    every axis plot post/.append style={
        every mark/.append style={scale=2}
    }
]
\addplot[color={rgb:red,0;green,157;blue,178}, mark=square, mark options={solid}]
table[x=para, y=tmall_ndcg] {latFactor.txt};
\legend{\large Tmall};
\end{axis}
\end{tikzpicture}

\begin{tikzpicture}
\begin{axis}[
    xlabel={Hidden State Dimensionality $d$},
    ylabel={NDCG@10},
    xmin=1,xmax=33,
    ymin=0.28,ymax=0.34,
    legend columns=1,
    legend cell align=right,
    grid=both,
    every axis plot/.append style={ultra thick},
    every tick label/.append style={scale=1.3},
    label style={scale=1.8},
    legend style={
        nodes={scale=1.5, transform shape},
        legend image post style={scale=1.5},
        },
    legend style={at={(1,0)},anchor=south east},
    every axis plot post/.append style={
        every mark/.append style={scale=2}
    }
]
\addplot[color={rgb:red,0;green,157;blue,178}, mark=square, mark options={solid}]
table[x=para, y=ijcai_ndcg] {latFactor.txt};
\legend{\large IJCAI};
\end{axis}
\end{tikzpicture}
    \end{adjustbox}
    \caption{Impact of hidden state dimensionality of \model framework on BeiBei, Tmall, IJCAI datasets.}
    \vspace{-0.1in}
    \label{fig:hyperparam}
\end{figure}
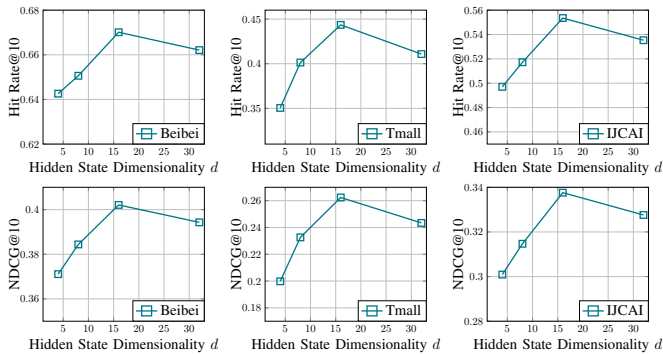

\begin{itemize}[leftmargin=*]

\item \textbf{Comparison with Different Hidden Dimensionality}. Our model results with different dimension size of hidden states are shown in Figure~\ref{fig:hyperparam}. We observe that larger embedding size does not always bring the positive effect for improving model performance, especially for sparse experimented datasets. The larger size of hidden state dimensionality may lead to the overfitting issue. We set the hidden dimensionality $d=16$ as the default value in our \emph{\model}. \\\vspace{-0.05in}


\item \textbf{Comparison with Different Graph Model Depth}.
To investigate the performance of our MBGNN method by stacking multiple graph neural layers, we conduct experiments by varying the number of graph-based embedding propagation layers. As shown in Table VII, we can observe that \emph{\model}-2 and \emph{\model}-3 obtain consistent improvement over \emph{\model}-1 which merely considers the first-order neighbors for message passing. We attribute the performance improvement to the encoding of collaborative relations based on our considered second- and third-order neighboring node dependency. With the further increase of model depth from two to three graph layers, the performance slight degrades with the configuration of deep graph neural architecture. The reason may lie in that deep graph neural framework tends to be overfitting and involve the over-smoothing issue in the generated user/item representations. According to the statistical information from our experimented Tmall data, with the consideration of three-hop connections, a large percentage of user-item pairs may be connected, which unavoidably leads to the over-smoothing issue of making user embeddings indistinguishable.\\\vspace{-0.05in}



\item \textbf{Comparison with Different Number of Channels}. We vary the number of embedding channels in our multi-channel behavior embedding layer. The results in terms of HR@10 and NGCD@10 are presented in Table~\ref{tab:memory_units}, from which we notice that the performance of \emph{\model} is improved at first, with the increase of behavior representation channels. But we can observe that the recommendation performance degrades with the further increase of channel numbers, due to the overfiting. Hence, behavior embedding channels with the dimension of 16 is enough for encoding interaction semantics.

\end{itemize}

\begin{table}[t]
    \caption{Computational cost (seconds) study.}
    \centering
    \small
    \vspace{-0.05in}
    \begin{tabular}{lccc}
         \toprule
         Models& BeiBei \ \ \  & Tmall  & IJCAI \\
         \midrule
         NADE & 4.1s & 26.9s & 60.4s\\
         CF-UIcA & 11.5s & 61.7s & 139.1s\\
         ST-GCN & 12.6s & 58.5s & 94.8s\\
         NGCF+M & 15.8s & 74.6s & 152.3s\\
         NMTR & 14.0s & 37.3s & 118.0s\\
         MBGCN & 17.4s & 85.3s & 186.5s\\
         MATN  & 11.5s & 74.7s & 196.5s\\
         DIPN & 53.2s & 172.6s & 284.6s\\
         \hline
         \emph{\model} & 14.3s & 58.6s & 101.1s \\
         \hline
    \end{tabular}
    \label{tab:time}
\end{table}

\subsection{Computational Cost Analysis (RQ6)}
\label{sec:time_eval}

Our evaluation also includes the computational cost investigation of our \emph{\model} model and several representative methods in terms of their implementation time on different datasets. We evaluate the computational cost of all compared methods on the machine of NVIDIA TITAN RTX GPU with the configurations of Intel Xeon W2133 CPU 3.6G Hz and 64GB RAM. For fair comparison, we apply the same setting of hidden state dimensionality for all methods. For graph-based methods, the number of embedding propagation layers is set as 2. From the reported evaluation results in Table~\ref{tab:time}, we can observe that our \emph{\model} model can achieve comparable model efficiency compared with other baselines in terms of the implementation time. In particular, when competing with multi-behavior recommendation baselines (NGCF+M, MBGCN), our \emph{\model} requires less implementation time, which indicates the efficiency of our multi-behavior graph neural framework. Additionally, compare with the autoregressive collaborative filtering model-CF-UIcA, our \emph{\model} can still achieve competitive model efficiency with the incorporation of multi-typed behaviour context. In summary, the above observations justify the scalability of our proposed \emph{\model} in dealing with large-scale user behavior data for recommendation.\\



\subsection{Model Interpretation with User Study (RQ7)}
\label{subsec:case}

\begin{figure}[t]
    \centering
    \includegraphics[width=0.93\columnwidth]{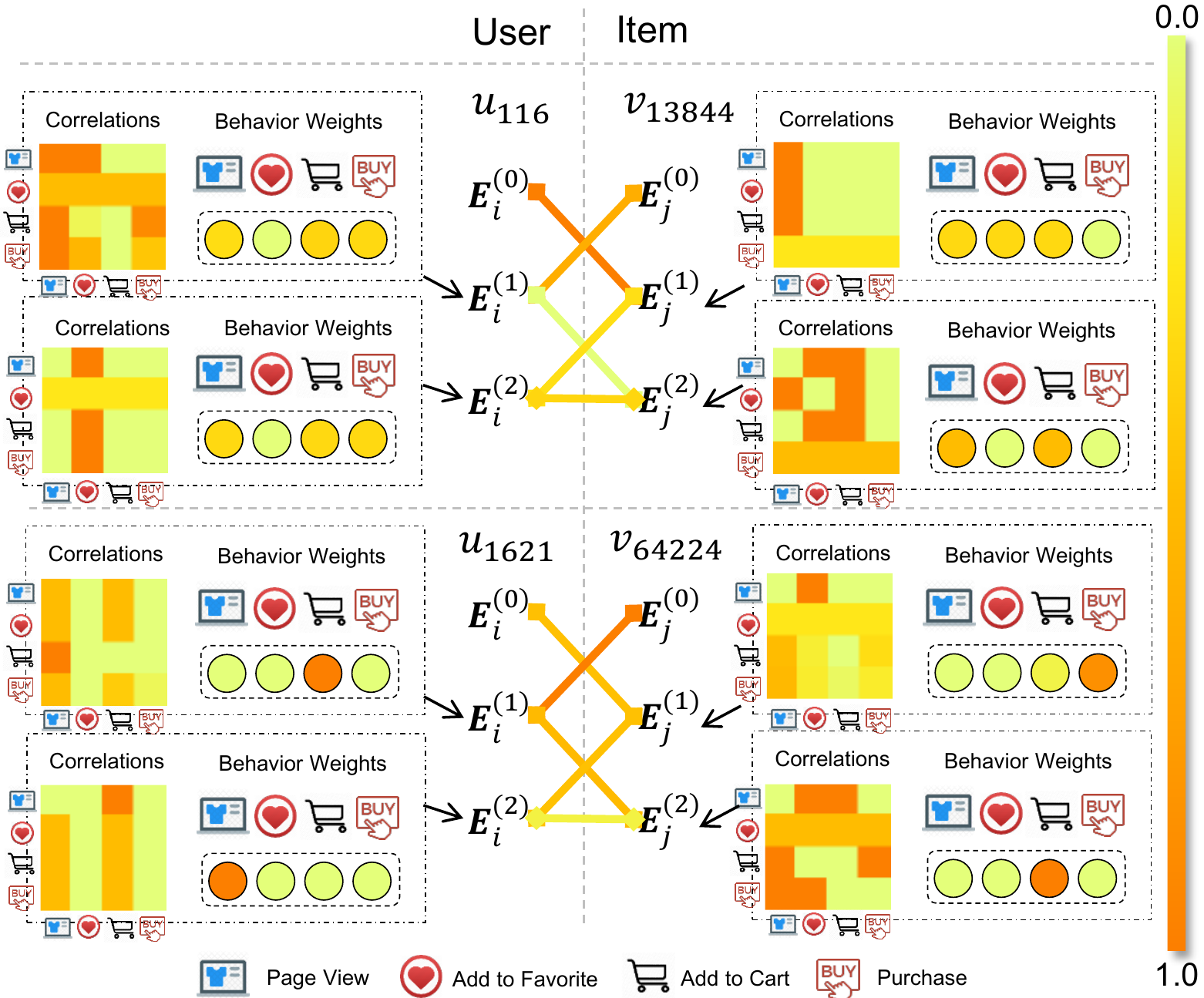}
    \caption{Interpretation study of multi-behavior inter-dependency in our \emph{\model} \wrt\ behavior inter-dependency modeling, behavior pattern aggregation and cross-layer mutual relation learning. Dark color indicates higher relevance. Best viewed in color.}
    \label{fig:case_study_weights}
\end{figure}

To analyze the multi-behavior dependency interpretation of our proposed \model\ framework, we conduct user studies with identified real user examples. We show the study results in Figure~\ref{fig:case_study_weights}. In this figure, the cross-type behavior dependencies between user ($u_{116}$, $u_{1621}$) and item ($v_{13844}$, $v_{64224}$) are shown with the learned quantitative dependency weights. Specifically, $E_i^{(1)}$ and $E_i^{(2)}$ denote our produced user representation encoded from the $1^{st}$ and $2^{nd}$ graph-based embedding propagation layer, respectively. Similarly, $E_j^{(1)}$ and $E_j^{(2)}$ represents the encoded item embeddings from the $1^{st}$ and $2^{nd}$ message passing layers, respectively. From the study results, we summarize the key observations as follows:


\begin{itemize}[leftmargin=*]

\item \textbf{Encoded Behavior Inter-Correlations}. In this user study, we present the learned behavior inter-correlation matrix with the dimension of $\mathbb{R}^{4\times 4}$ to reflect the pairwise correlations between different types of user behaviors, \ie, page view, add-to-cart, tag-as-favorite and purchase.\\\vspace{-0.1in}



\item \textbf{Type-specific Behavior Pattern Fusion}. In our \model\ recommendation framework, we design the multi-behavior pattern aggregation module with the aim of integrating type-specific behaviour patterns for making final recommendation. In particular, each user is associated with a learned attention-based behavior importance vector with the dimension of $\mathbb{R}^{1\times 4}$ (as shown in Figure~\ref{fig:case_study_weights}). For example, we can observe that users who view item $v_{13844}$ are more likely to purchase it compared with item $v_{64224}$.\\\vspace{-0.1in}



\item \textbf{Cross-layer Mutual Relation Encoding}. In our multi-layer graph neural framework, we introduce a mutual relation encoding component to explicitly aggregate representations from different hops in the multi-behavior interaction graph. $E_i^{(1)}$ and $E_i^{(2)}$ represents the encoded embeddings of user $u_i$ from his/her first- and second-order neighboring nodes. In Figure~~\ref{fig:case_study_weights}, the correlations among hop-aware user/item representations are shown with different connection lines. From the visualization results, We can observe that cross-layer user/item embeddings (\eg, $E_i^{(0)}$ and $E_j^{(1)}$) are often highly correlated with each other compared with the embeddings of the same layer (\eg, $E_i^{(2)}$ and $E_j^{(2)}$).


\end{itemize}



\subsection{Effect of Graph Sampling}
In this section, we investigate the effect of our graph sampling algorithm on the model performance by testing the prediction accuracy of \model\ with different number of training and testing sub-graphs. In specific, \model\ is trained with sub-graphs containing 5000, 10000, 20000, 40000 nodes, and is tested using input sub-graphs containing 5000, 10000, 20000, 40000, 60000 nodes. The results are shown in Table~\ref{tab:graphSampling}, from which we can conclude that testing on larger sub-graphs always yields better performance, while training on larger sub-graphs does not always result in better performance. This is because training with smaller sub-graphs may serve as regularization operation for predictions.

\begin{table}[t]
    \scriptsize
    \centering
	\setlength{\tabcolsep}{0.7mm}
    \caption{Influence of the sub-graph sampling scale.}
    \label{tab:graphSampling}
    \begin{tabular}{|c|c|c|c|c|c|c|c|c|c|c|}
        \hline
        \multirow{3}{*}{Train. $N$} & \multicolumn{10}{c|}{Number of Sub-graph Size $N$ When Testing}\\
        \cline{2-11}
        & \multicolumn{2}{c|}{5,000} & \multicolumn{2}{c|}{10,000} & \multicolumn{2}{c|}{20,000} & \multicolumn{2}{c|}{40,000} & \multicolumn{2}{c|}{60,000}\\
        \hline
        & HR & NDCG & HR & NDCG & HR & NDCG & HR & NDCG & HR & NDCG\\
        \hline
        \hline
        \multirow{1}{*}{5,000} & 0.365 & 0.201 & 0.409 & 0.230 & 0.463 & 0.266 & 0.527 & 0.310 & 0.552 & 0.338\\
        \hline
        \multirow{1}{*}{10,000} & 0.359	& 0.198 & 0.407 & 0.229 & 0.464 & 0.266 & 0.529 & 0.307 & 0.552 & 0.336\\
        \hline
        \multirow{1}{*}{20,000} & 0.357 & 0.196 & 0.407 & 0.228 & 0.466 & 0.270 & 0.537 & 0.326 & 0.554 & 0.338\\
        \hline
        \multirow{1}{*}{40,000} & 0.322 & 0.168 & 0.367 & 0.197 & 0.424 & 0.236 & 0.500 & 0.292 & 0.548 & 0.330 \\
        \hline
    \end{tabular}
\end{table}
\section{Related Work}
\label{sec:relate}

\subsection{Neural Network Collaborative Filtering Models}
Collaborative Filtering (CF) has become one of the most important paradigms for personalized recommender systems in real-life platforms~\cite{wang2019unified,huang2021neural}. The general idea of CF models is that users may share similar preference if they interact with similar items~\cite{hu2008collaborative}. In recent years, many efforts have been made to augment the CF techniques with deep neural network models~\cite{shi2019deep}. These methods apply different neural mechanisms (\eg, autoencoder, attention mechanism, graph neural network) in the matching function to parameterize users and items into latent representation space. The learned representations of users and items can be used to estimate the likelihood of unobserved interactions.

Some of studies follow this research line to enable the non-linear feature interactions with the multi-Layer feed-forward network, such as NCF~\cite{he2017neuralncf} and DMF~\cite{xue2017deep}. To consider item relational data into the CF model, the relational collaborative filtering (RCF~\cite{xin2019relational}) framework designs neural two-stage attention mechanism to enhance the item embedding process. Another recent research line of recommendation models is to explore the user-item interaction graph to capture the collaborative filtering signals. For example, NGCF~\cite{wang2019neural} is developed based on the high-hop information propagation framework to guide the user/item representation procedure. ST-GCN~\cite{zhang2019star} is another graph learning model to encode user-item interaction patterns with an encoder-decoder framework. In addition, to bridge the logical reasoning and representation learning in recommender systems, a neural collaborative reasoning approach (NLR)~\cite{chen2020neural} is proposed to incorporate the logic priors into the neural architecture.


\subsection{Recommendation with Multi-Behavior Modeling}
There exist some research works aiming at enhancing recommendation models by considering multi-typed behavior of users~\cite{tang2016empirical}. In those methods, the implicit user-item feedback from auxiliary behaviors (\eg, click, add-to-cart) are considered as behavior contextual signals to predict target user behaviors (\eg, purchase)~\cite{jin2020multi,xia2021graph}. For example, multi-task learning frameworks are developed to perform the joint training among the prediction tasks of different behavior types~\cite{gao2019neural}. However, those methods reply on the predefined dependent relationships between different types of user behaviors, and can hardly be reflective of the complex multi-behaviour context in practical scenarios.

To capture the correlations between different types of behaviors, MATN~\cite{matnsigir20} utilizes the attention network for multi-behavior information aggregation. Both browsing and buying behaviors of users are considered in DIPN~\cite{guo2019buying} with an attention-based RNN model.
However, the high-order behavior dependent structures have been overlooked in them. In this work, the proposed \model\ framework aims to encode the high-order collaborative signals in the embedding function. Additionally, graph-based methods have been designed to tackle the multi-behavior recommendation problem. Specifically, Zhang~\etal~\cite{zhang2020multiplex} employs the multiplex network embedding technique to generate behavior-aware embeddings. MBGCN~\etal~\cite{jin2020multi} is built on the graph convolutional network to propagate the behavior embeddings over the interaction graph. Our new \model\ differs from those graph-based models from two perspectives: i) we discriminate the influence between various behaviour patterns through a dual-stage relation learning scheme. The designed new message passing paradigm endows the multi-behavior graph neural network with the capability of encoding behavior-aware characteristics and dependencies simultaneously. ii) The high-order collaborative signals are aggregated across different graph layers explicitly, under the cross-layer message passing architecture.

\subsection{Graph Neural Networks for Recommendation}
In view of the effectiveness of Graph Neural Networks (GNNs), GNNs have been widely used to perform the representation learning over the graph-structured data~\cite{wang2020gcn,wu2020comprehensive,liu2021item,chen2020handling,li2021learning,tang2019coherence,kipf2016semi}. 
Recent research works apply the graph neural network to model user-item interactions in recommender systems: PinSage~\cite{ying2018graph} is a graph convolutional network to propagate embeddings over the pin-board bipartite graph structure. Additionally, modeling the dynamic user-item interactions has attracted much attention for recommender systems~\cite{chen2021temporal,yan2019cosrec}. To encode the sequential patterns, graph neural networks have been utilized to consider the transitions between items of session sequences in SRGNN~\cite{wu2019session} and MTD~\cite{2021graph}, or user interaction sequence in H2SeqRec~\cite{li2021hyperbolic}. In addition, the graph diffusion network~\cite{wu2020diffnet} and graph attention mechanism~\cite{wu2019dual} have been utilized to capture the influence among users, so as to incorporate the social relations into the recommendation and alleviate the data sparsity issue.


\section{Conclusion}
\label{sec:conclusion}

In this work, we contribute a new end-to-end framework (\model) for multi-behavior recommendation via the modeling of cross-behavior inter-dependencies under a high-order graph learning architecture. In our \model\ model, we first learn the dependent relationships among various types of user interactions with a behavior-aware message passing mechanism. Additionally, a designed high-order mutual relation learning scheme is integrated with the graph neural architecture, so as to encode the implicit dependencies between layer-specific behavior representations. When evaluated on three real-world datasets, our framework achieves significantly better recommendation performance as compared to various baselines. Further studies on model ablation show the rationality of designed key components in our proposed recommendation framework. In future, we would like to integrate the causal effect analysis~\cite{bonner2018causal} with our designed multi-behavior graph neural paradigm, in order to infer the causal relations from observed user behaviors and identify the implicit factors which influence user preference.



\section*{Acknowledgments}
We thank the reviewers for their valuable feedback and comments. This research work is supported by the research grants from the Department of Computer Science \& Musketeers Foundation Institute of Data Science at the University of Hong Kong (HKU). The research is also partially supported by National Nature Science Foundation of China (62072188), Major Project of National Social Science Foundation of China (18ZDA062), Science and Technology Program of Guangdong Province (2019A050510010).

\bibliographystyle{abbrv}
\bibliography{refs}

\begin{IEEEbiography}[{\includegraphics[width=1in,height=1.05in,clip,keepaspectratio]{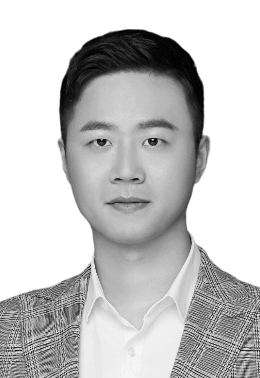}}]{Lianghao Xia}
	is currently a postdoctoral fellow in the Department of Computer Science \& Musketeers Foundation Institute of Data Science, at the University of Hong Kong. He received his B.E. and PhD degrees from South China University of Technology in 2017 and 2021, respectively. His research interests include data mining, graph neural networks and recommender systems. His research work has appeared in several major international conferences and journals such as SIGIR, AAAI, IJCAI, ICDE, CIKM, ICDM as well as ACM TOIS.
\end{IEEEbiography}\vspace{-0.3in}

\begin{IEEEbiography}[{\includegraphics[width=1in,height=1.15in,clip,keepaspectratio]{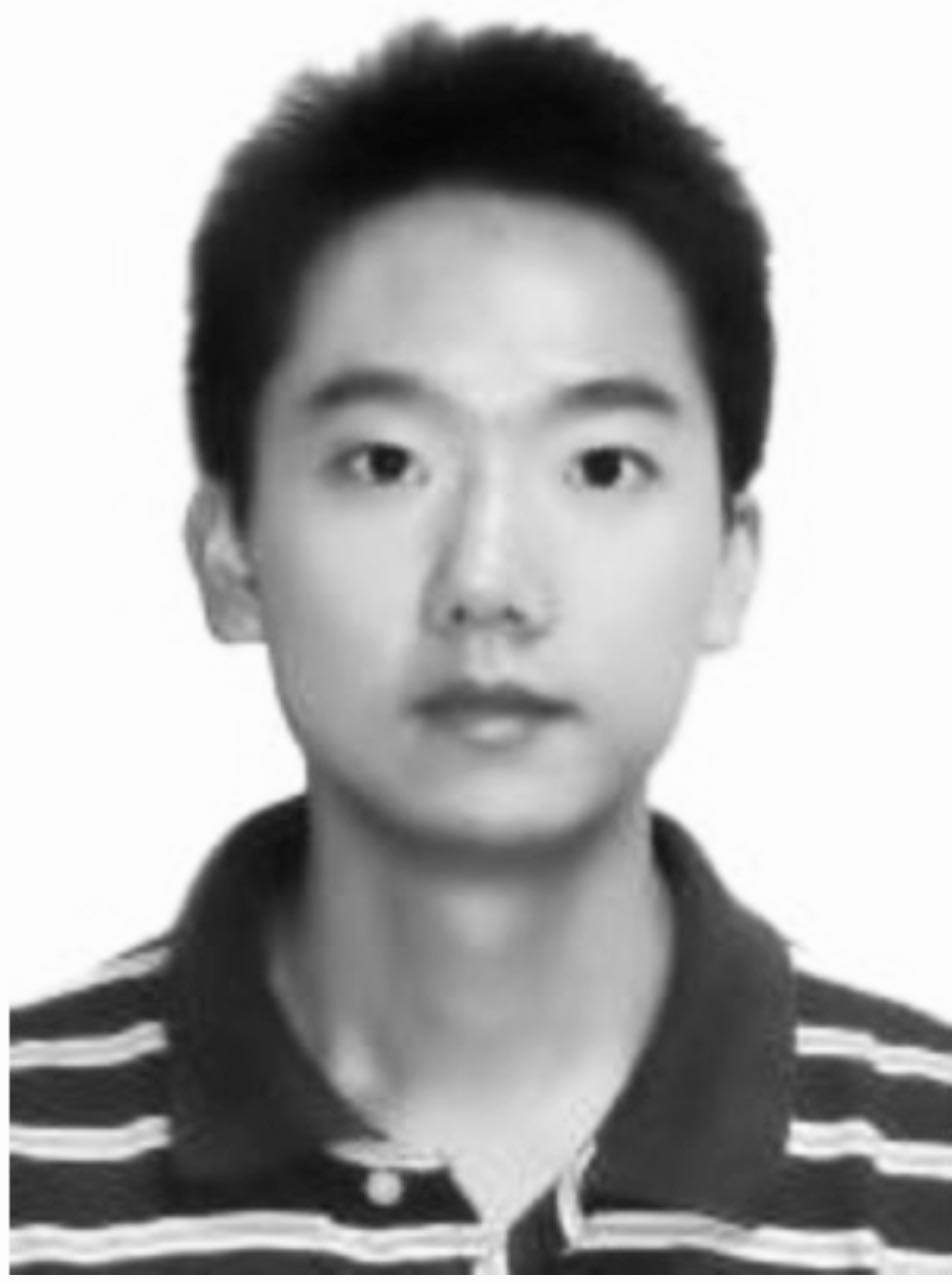}}]{Chao Huang}
is a tenure-track assistant professor in the Department of Computer Science \& Musketeers Foundation Institute of Data Science, at the University of Hong Kong. He obtained the PhD degree from the University of Notre Dame in 2019. His research focuses on applied machine learning, graph neural networks, recommendation and spatial-temporal data mining. His work has appeared in several major international conferences such as KDD, WWW, SIGIR, IJCAI, AAAI, WSDM and etc. He has served as the PC member for several top conferences including KDD, WWW, SIGIR, WSDM, AAAI, IJCAI, NIPS, ICLR and etc. Additionally, he has been recognized as the outstanding reviewer in both ACM WSDM'2020 and WSDM'2022 conference.
\end{IEEEbiography}\vspace{-0.3in}

\begin{IEEEbiography}[{\includegraphics[width=1in,height=1.15in,clip,keepaspectratio]{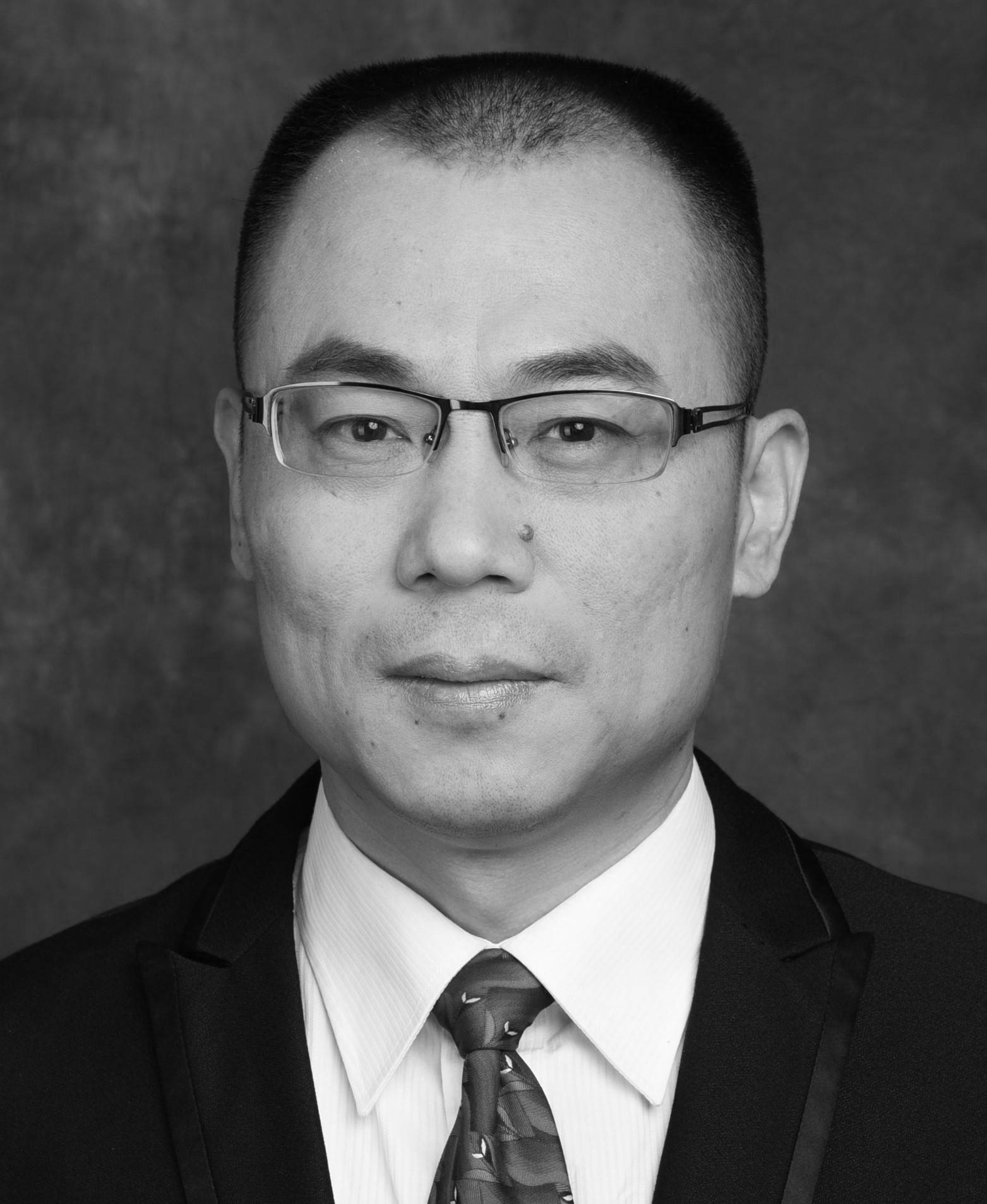}}]{Yong Xu}
is a Professor at the School of Computer Science and Engineering in South China University of Technology. His research interests include machine learning, pattern recognition and big data analysis. He has published over 80 research papers in refereed journals and conferences (\eg, SIGIR, AAAI, IJCAI, CIKM, CVPR, NIPS, ICCV, TIP, TMM and TOIS) and been serving as PC for conferences \& journals including AAAI, CVPR, ICCV, TIP and etc. Dr. Xu is a member of the IEEE Computer Society and the ACM.
\end{IEEEbiography}\vspace{-0.3in}

\begin{IEEEbiography}[{\includegraphics[width=1in,height=1.15in,clip,keepaspectratio]{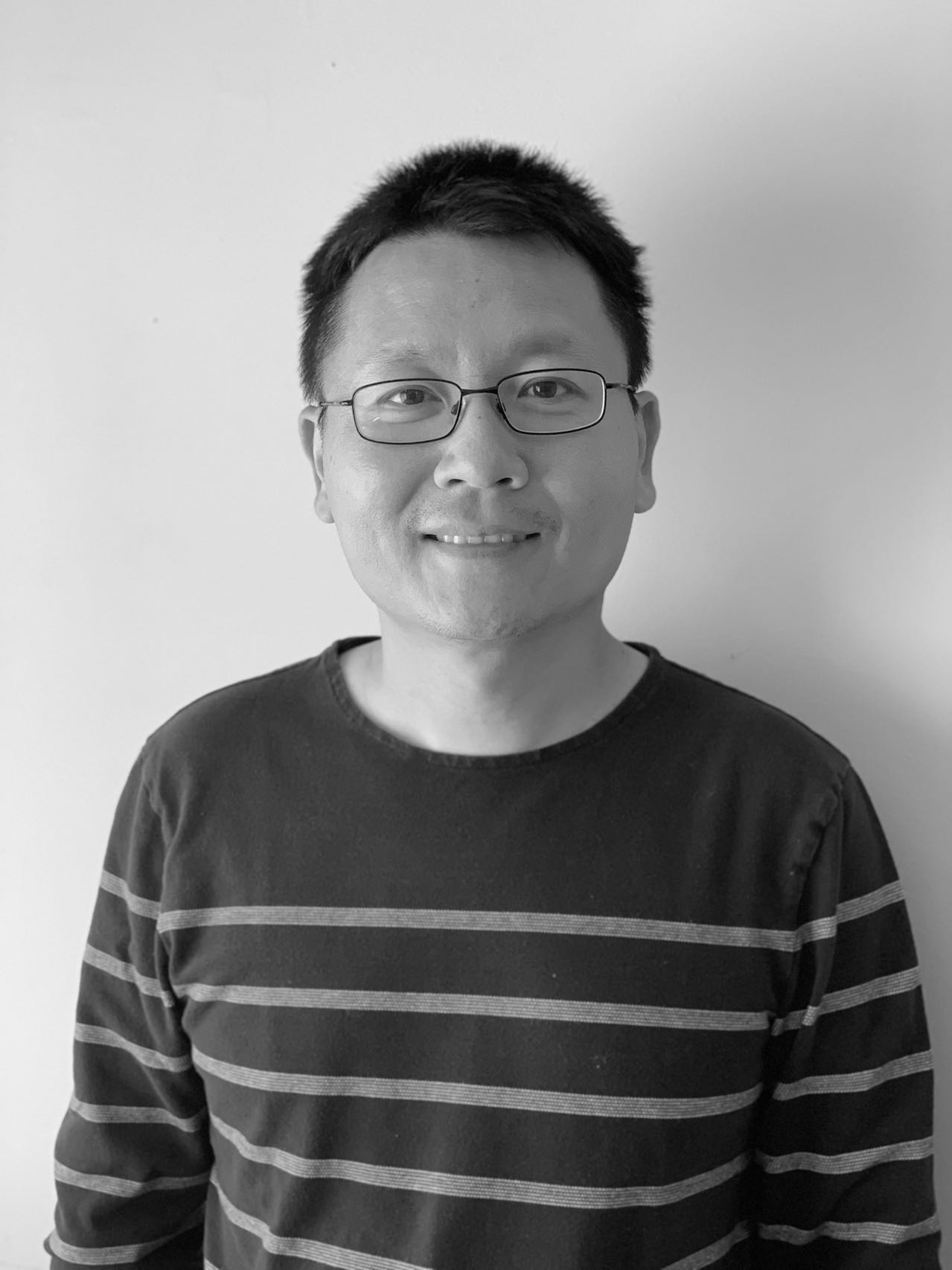}}]{Peng Dai}
is a principal scientist at JD silicon valley research center. He obtained his PhD degree from the University of Washington in 2011. His research interests include artificial intelligence and machine learning. He has published broadly in top conferences and journals, such as AAAI, IJCAI, ICAPS, SIGIR, CIKM, WWW, CSCW, JAIR, AIJ, etc. His dissertation won Honorable Mention of 2012 ICAPS Best Dissertation Award.
\end{IEEEbiography}\vspace{-0.2in}



\begin{IEEEbiography}[{\includegraphics[width=1in,height=1.15in,clip,keepaspectratio]{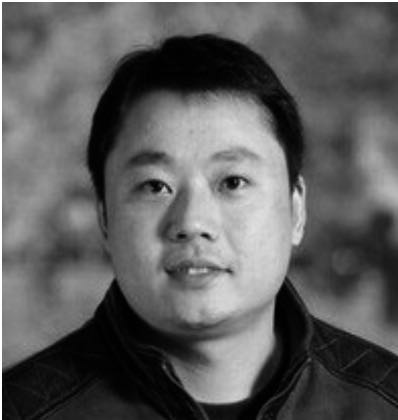}}]{Liefeng Bo}
is a vice president and distinguished scientist at JD silicon valley research center. His research interests includes deep learning, computer vision, and big data systems. He held an affiliate faculty at the University of Washington. His work has appeared in several major international conferences, such as CVPR, NIPS, ICML, IJCAI, AAAI, ICRA. He has served as the PC member for several top conferences including NIPS, CVPR, ICCV, ECCV, TPAMI, TIP and etc. His paper has won the best vision paper award in ICRA 2011 and been selected into finalist for best vision paper award in ICRA 2014.
\end{IEEEbiography}

\ifCLASSOPTIONcaptionsoff
  \newpage
\fi

\end{document}